\documentclass{article}

\usepackage{microtype}
\usepackage{graphicx}
\usepackage{subcaption}
\usepackage{accessibility}
\usepackage{booktabs} % for professional tables
\usepackage{tcolorbox}
\usepackage{amssymb}
\usepackage{caption}  
\usepackage{pifont} 
\usepackage{siunitx}
\usepackage{algpseudocode}
\usepackage{listings}
\usepackage{hyperref}
\usepackage{amsmath} 
\usepackage{ragged2e}
\usepackage{array}
\usepackage{pifont}
\usepackage{booktabs, tabularx, array}
\usepackage{pifont}
\usepackage{threeparttable,threeparttablex}
\usepackage{pifont}
\usepackage{booktabs}
\usepackage{pifont}
\newcommand{\cmark}{\ding{51}}% check mark
\newcommand{\xmark}{\ding{55}}% cross mark
\usepackage{accessibility}
\usepackage[accepted]{mlsys2025}

\usepackage{siunitx}
\newcommand{\tick}{\checkmark}   % or however you define the symbol

\usepackage{enumitem}

\mlsystitlerunning{SONIQ: System-Optimized Noise-Injected Ultra-Low-Precision Quantization with Full-Precision Parity}

\begin{document}

\twocolumn[
\mlsystitle{SONIQ: System-Optimized Noise-Injected Ultra-Low-Precision Quantization with Full-Precision Parity}

% unless you've provided the [accepted] option to the mlsys2025


\mlsyssetsymbol{equal}{*}

\begin{mlsysauthorlist}
\mlsysauthor{Cyrus Zhou}{stanford}
\mlsysauthor{Pedro Savarese}{ttic}
\mlsysauthor{Zack Hassman}{uchicago}
\mlsysauthor{Vaughn Richard}{uchicago}
\mlsysauthor{Michael DiBrino}{futurewei}
\mlsysauthor{Michael Maire}{uchicago}
\mlsysauthor{Yanjing Li}{uchicago}
\end{mlsysauthorlist}

\mlsysaffiliation{stanford}{Department of Computer Science, Stanford University, CA, USA}
\mlsysaffiliation{uchicago}{Department of Computer Science, University of Chicago, Chicago, IL, USA}
\mlsysaffiliation{futurewei}{FutureWei Technologies, Austin, TX, USA}
\mlsysaffiliation{ttic}{TTI-Chicago, Chicago, IL, USA}

% \mlsyscorrespondingauthor{Cyrus Zhou}{zhouzk@uchicago.edu}
\mlsyscorrespondingauthor{Yanjing Li}{yanjingl@uchicago.edu}
% \mlsyscorrespondingauthor{Eee Pppp}{ep@eden.co.uk}

\mlsyskeywords{Machine Learning, MLSys}

\vskip 0.3in

\begin{abstract}
Ultra-low-precision inference can sharply reduce memory and latency but often degrades accuracy and relies on specialized hardware. We present SONIQ, a system-optimized, noise-injected quantization framework that learns per-channel mixed precision for both weights and activations while training under the same rules used at inference. By injecting hardware-calibrated quantization noise during training, SONIQ steers models toward the discrete arithmetic used at deployment—without bespoke runtimes. Across CNNs and Transformers, SONIQ achieves up to 16× and 7× compression, respectively, while matching or exceeding full-precision accuracy. Measured end-to-end, SONIQ delivers up to 7.3× CPU speedup over strong INT8 baselines and up to 6.3× (vector units) / 2.8× (tensor cores) GPU speedup relative to FP16. A practical outcome is that two per-channel precision levels—one in the 1–4-bit range and one in the 4–8-bit range—suffice in practice; at inference, each channel selects one of the two, keeping kernels simple and fast. To our knowledge, SONIQ is the first framework to reach or surpass full-precision accuracy under ultra-low (1–4 bits per parameter) regimes while remaining deployable on commodity hardware, narrowing the gap between quantization theory and practical, high-throughput inference.
\end{abstract}
]



\printAffiliationsAndNotice{}  % leave blank if no need to mention equal contribution
\definecolor{symbcolor}{RGB}{204, 0, 0}
\newcommand{\cyrus}[1]{{\color{blue}[CZ: #1]}}

\definecolor{darkergreen}{RGB}{0,100,0}
\newcommand{\zack}[1]{{\color{darkergreen}[ZH: #1]}}
\newcommand{\llama}[1]{{\color{brown}[llama: #1]}}

\newcommand{\symb}[1]{\boldsymbol{\color{symbcolor}#1}}
\newcommand{\vaughn}[1]{{\color{green}[VR: #1]}}
\newcommand{\michael}[1]{{\color{teal}[MD: #1]}}
\newcommand{\li}[1]{{\color{orange}[li: #1]}}
\newcommand{\qz}[1]{{\color{purple}[QZ: #1]}}

\newcommand{\todo}[1]{\color{red}{\textbf{TODO: #1}}}
\newcommand{\RNum}[1]{\uppercase\expandafter{\romannumeral #1\relax}}
\newcommand{\yl}[1]{{\color{purple}#1}}

\def\FGMPNN{SoniqNet}
\def\FWName{SONIQ}
\def\MACName{SONIQ MAC Unit}
\def\FWMeaning{\textbf{N}oise-\textbf{E}nhanced \textbf{P}recision \textbf{T}raining for \textbf{U}ltra-low-bit \textbf{N}etwork \textbf{E}fficiency}

\definecolor{darkgreen}{rgb}{0.0, 0.5, 0.0} 
\definecolor{darkred}{rgb}{0.5, 0.0, 0.0}
\def\YesGreen{\textcolor{darkgreen}{Yes}}

\def\NoRed{\textcolor{darkred}{No}}
% Custom macros

\newtcolorbox{highlighted}{
  colback=yellow!10,
  colframe=yellow!50!black,
  boxrule=0.5pt,
  arc=1mm,
  top=0mm,
  bottom=0mm,
  left=0mm,
  right=0mm,
  boxsep=0mm,
  toprule=0pt,
  bottomrule=0pt
}

\lstset{
  basicstyle=\scriptsize\ttfamily, % set the font size and monospace font
  escapeinside={(*@}{@*)},
}
\lstdefinestyle{customverilog}{
    language=Verilog,
    basicstyle=\ttfamily\footnotesize,
    keywordstyle=\color{blue},
    identifierstyle=\color{black},
    commentstyle=\color{green},
    numberstyle=\tiny\color{gray},
    stringstyle=\color{red},
    breaklines=true,
    showstringspaces=false,
    tabsize=3,
    frame=single
}
\section{Introduction} \label{sec:intro}

% Memory and compute efficiency are both pivotal to real-world applications of deep learning, which have exceeded or matched human-level performance on numerous tasks \cite{yang2024harnessing,street2024llms,zhao2023more}. Deploying deep learning models on edge devices and smartphones where memory resources and computational power are limited have gathered more and more interest \cite{zhang2018shufflenet,chen2019eyeriss,lin2024awq}. In numerous applications including autonomous vehicles and real-time translation, optimizing neural network efficiency is crucial to facilitate deployment on these resource-constrained devices. In large-scale deployments, such as data centers serving millions of users, computational resources and energy consumption become substantial expenses \cite{zhao2023recd,hazelwood2018applied}. Optimizing models in these scenarios can substantially lower costs while maintaining performance and leading to more sustainable operations \cite{nguyen2024s,mcdonald2022great,zhu2024nanoflow}.

Memory and compute efficiency are pivotal to the real-world applications of deep learning, which has matched or exceeded human-level performance across numerous tasks \cite{yang2024harnessing,street2024llms,zhao2023more}. As interest grows in deploying deep learning models on edge devices and smartphones—where memory and computational resources are highly constrained—optimizing neural network efficiency has become increasingly critical \cite{zhang2018shufflenet,chen2019eyeriss,lin2024awq}. Similarly, large-scale deployments, such as data centers, face significant costs associated with computational resources and energy consumption \cite{zhao2023recd,hazelwood2018applied}, motivating model optimizations that improve both cost and sustainability \cite{nguyen2024s,mcdonald2022great,zhu2024nanoflow}.

\vspace{0.5em}

Inference is a prominent bottleneck in modern machine-learning systems. Production services already issue tens of trillions of model queries per day, and at hyperscaler scale, the cost of serving those queries can exceed 90\% of the total ML infrastructure budget \cite{romero2021infaas}. User-facing applications further impose stringent service-level objectives—e.g., delivering answers within $\approx$ 100 ms end-to-end—because even small latency regressions degrade engagement and may jeopardize safety-critical deployments \cite{nigade2024inference}. Moreover, the continuous stream of forward passes dominates the operational energy of AI, accounting for approximately 60\% of the ML electricity footprint in large data centers \cite{patterson2024energy}. At the same time, techniques for improving inference efficiency, such as aggressive quantization or pruning, not only degrade accuracies but have also been shown to amplify error on minority sub-populations and exacerbate algorithmic bias \cite{hooker2020characterising}. These observations underscore the importance of co-optimizing latency, energy, and accuracy without compromising statistical fidelity and fairness.  

Running deep models on edge devices presents additional challenges, requiring the simultaneous achievement of three goals: (1) low latency (e.g., sub-tens-of-millisecond), (2) minimal memory footprint (e.g., kB-scale), and (3) uncompromised accuracy.  Interactive mobile and AR/VR experiences collapse if motion-to-photon latency exceeds $\sim$20 ms, demanding sub-tens-of-millisecond inference even on phone-class CPUs/NPUs~\cite{lin2017latency}.  Memory budgets are unforgiving: microcontroller deployments must fit \emph{all} weights, activations, and runtime into at most 256 kB of SRAM and a few MB of flash~\cite{lin2022ondevice}.  Accuracy, however, cannot be compromised. In smartphone-based atrial fibrillation screening, false negatives can lead to missed strokes~\cite{barbosa2025af}. Similar strictness applies to safety-critical perception systems: even a few-percentage-point drop in obstacle detection recall on nano-drones significantly increases collision rates during 25–40 fps navigation loops~\cite{navardi2022drone}. % These examples illustrate an edge‐computing regime where \emph{sub-tens-of-millisecond efficiency} and \emph{kB-scale memory} must be delivered \emph{without} any degradation in task fidelity, 
These requirements motivate techniques—such as ours—that jointly optimize all three dimensions.

Quantization \cite{lin2024awq,liang2021pruning,liu2023ultra,tang2022mixed,zhong2022dynamic} is among the most widely adopted techniques for reducing model size and inference latency while maintaining competitive accuracy. State-of-the-art methods achieve high compression ratios through fine-grained precision mixing within individual network layers, applied either during training \cite{savarese2022not,wang2022learnable,tang2022mixed} or via post-training optimizations \cite{chauhan2023post,zhao2024atom,zhou2025lowra,dong2023emq}. These approaches enable aggressive quantization, even allowing each network parameter/activation to adopt an individually optimized precision, achieving compression while preserving or exceeding full-precision accuracy.

Despite this progress, there is a gap between \emph{algorithmic}
quantization goals and \emph{system-level} deployment realities.
Several state-of-the-art approaches adopt precision patterns
(e.g., 3-bit activations, per-parameter scales, or irregular group
sizes) that lack mature support in commodity GPU, CPU, and accelerators, so model size reduction does not always
translate into lower wall-clock latency or power consumption
\cite{wang2022learnable,guo2023olive}.
Conversely, hardware-friendly schemes that use uniform 2- or 4-bit formats can achieve inference speed targets, but often sacrifice a noticeable amount of accuracy~\cite{yang2021bsq,choi2019accurate}.
Furthermore, many recent proposals are tuned for a single
application domain—language models 
\cite{lin2024awq,xiao2023smoothquant,zhao2023atom} or diffusion
generators~\cite{croitoru2023diffusion}, raising questions about their \emph{generality}.

\begin{figure*}[ht]
    \centering
    \includegraphics[width=0.9\linewidth]{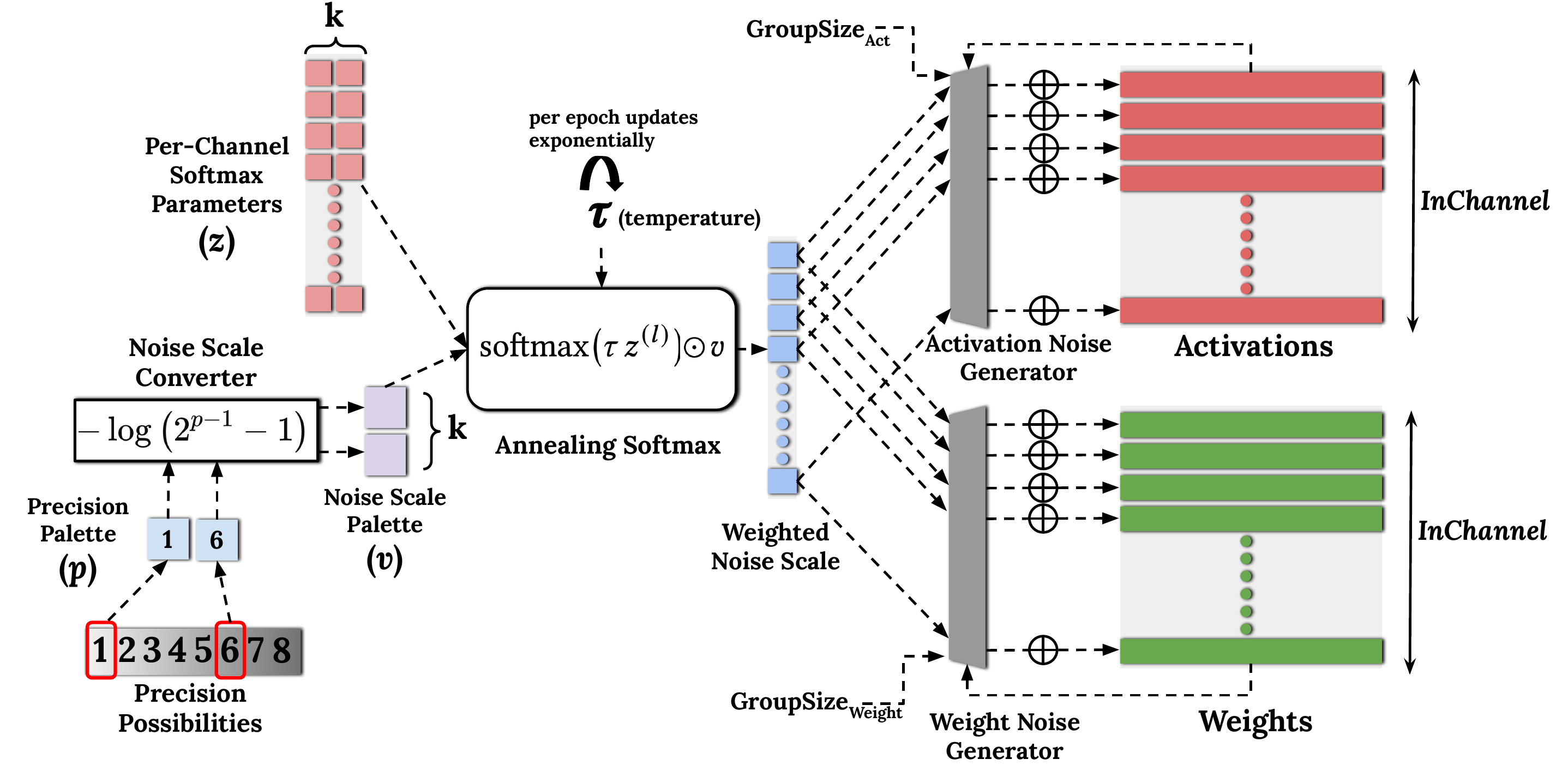}
    \caption{End-to-end \FWName{-QAT} workflow.  From left to right: (1) enumerate the bit-widths natively supported by the target accelerator and assemble them into a \emph{precision palette}; (2) use a temperature-annealed soft-assignment to choose a palette entry for every channel; (3) quantize weights and activations with hardware-calibrated noise drawn from the selected precisions, allowing joint optimization of parameters and bit-widths during back-propagation.}
    % \vspace{-1em}

    % \Description{Overview diagram of the QAT algorithm workflow, showing steps from precision palette formation to noise addition.}
    \label{fig:titan_qat}
    % \vspace{-0.7em}
    \vspace{0.5em}
\end{figure*}

The above observations motivate our work, \FWName{}, a \emph{system-aware} quantization approach
that \emph{jointly} optimizes compression, accuracy, and deployability
across diverse tasks and hardware. \FWName{} incorporates system-level inference costs into the training process. 
The \textbf{core innovation} of \FWName{} is \textit{\FWName{-QAT}} (\S\ref{sec:titan-qat}), a noise-injection-based quantization-aware training (QAT) algorithm that formulates system-aware quantization as a perturbation-bounded optimization problem. The overview of the \FWName{-QAT} algorithm is shown in Fig ~\ref{fig:titan_qat}, and the key features include:

\vspace{-1.2em}

\begin{itemize}[leftmargin=*]
\item \textbf{System Awareness during Training:}
During training we add group-wise, hardware-calibrated noise that mimics the rounding, saturation, and dynamic-range limits of the deployment engine. Perturbation magnitudes are clipped to each bit-width supported by a given hardware device, enabling the optimizer to roam the joint space of precisions and weights. The number of precision levels are also softly constrained. Thus, the model learns a mixed-precision configuration that achieves optimal accuracy under the device's latency, energy, and memory budgets.
\vspace{-0.2em}
\item \textbf{Fine-Grained Precision Learning:}  
Our approach learns per-channel mixed precisions for weights and activations, granting higher bit-widths to harder-to-quantize channels while compressing the remainder. We incorporate normalization-aware noise scaling to boost robustness and enhance quantization fidelity.
\vspace{-0.2em}
\item \textbf{Joint Weight-Activation Quantization:}  
Our approach enables joint mixed-precision quantization of both weights and activations within a unified training framework, maximizing compression for both. In contrast, state-of-the-art quantization techniques typically focus on either weights or activations, but not both simultaneously.  
\vspace{-0.2em}
\item \textbf{General and Efficient Deployment:}  
\FWName{} achieves up to 16× compression for CNNs and 7× for transformers. Moreover, The resulting models require only two precision levels for efficient and accurate inference, making them compatible with existing hardware architectures with minimal modifications. 
\vspace{-0.2em}
\item \textbf{End-to-End Performance Gains:}  
We integrate kernel-level optimizations into the framework to maximize inference efficiency, achieving up to 7.3× speedup for CNNs on CPU vector units against INT8 baselines and up to 6.3× and 2.8× speedups for transformers on GPU vector units and tensor cores against FP16 baselines, respectively, while preserving, or even exceeding, the accuracy of full precision neural networks.
\end{itemize}

To the best of our knowledge, \textbf{\FWName{} is the first system-aware quantization framework to surpass full-precision accuracy on \textit{both} convolutional and transformer neural network architectures}, even under challenging joint weight-activation quantization. Our results also reveal new insights into the key characteristics of optimized mixed-precision quantized networks.

This paper is organized as follows.
\S\ref{sec:background} reviews neural-network quantization, noise-injection training, and the system constraints that enable efficient low-bit inference.  
\S\ref{sec:titan-qat} introduces the \FWName{}-QAT algorithm.  
\S\ref{sec:training_setup}--\S\ref{sec:training_eval} detail the accuracy–compression study, and \S\ref{sec:inference_setup}--\S\ref{sec:inference-eval} report the corresponding inference results.  
Conclusions follow in \S\ref{sec:conclusion}. \textbf{Appendix \ref{app:novelty} clarifies the novelty and positioning of this work against related works}. Appendix \ref{app:qat_algo} shows the full algorithm of \FWName{-QAT}.
Appendix \ref{app:arch} describes our custom MAC unit for \FGMPNN{s} (i.e., networks trained with \FWName{-QAT}) and its architectural integration.

\section{Background}\label{sec:background}

\noindent
We set the stage for our method by
(i) assessing the essentials of neural network quantization (\S\,2.1) and noise-aware training (\S\,2.2);
(ii) unifying these threads by casting system-aware quantization as a structured noise-injection process that smooths the loss landscape (\S\,2.3); and
(iii) summarizing system-level constraints that any deployable low-precision model must satisfy (\S\,2.4).

\vspace{-0.2em}
\subsection{Neural Network Quantization} 
\textit{Quantization} compresses a network by storing and computing weights and activations at lower-precision bit-widths (e.g., 8 bits or less) instead of full 32-bit/16-bit floats.

Quantization of neural networks can be performed post-training or during training. Post-training quantization (PTQ) techniques have straightforward implementations and are cost-effective. However, these methods often result in performance degradation and are typically optimized for specific workloads only \cite{guo2023olive,zhao2024atom,lin2024awq}. 
%This limitation highlights a misalignment with the current cost distribution, where the sustained expenses of model serving necessitate more robust and adaptable optimization strategies. To better align with these shifting cost dynamics, 
Given these limitations, quantization-aware training (QAT) has emerged as a promising alternative. QAT integrates quantization into the training process, enabling models to maintain high accuracy and statistical fidelity. QAT can also target the more resource-intensive parts of inference to improve efficiency and scalability \cite{brown2024large,oliaro2024suffixdecoding}. Therefore, we focus on QAT.
\vspace{-0.5em}
\subsection{Noise-Aware Training}
Noise-aware training perturbs inputs, activations, or weights during optimization to smooth the loss landscape, regularize parameters, and harden models against distribution shift.  Classic theory shows that additive Gaussian noise acts as a Tikhonov penalty, which explains its generalization benefit~\cite{bishop1995training}.  The same idea underlies modern practice: \emph{dropout} randomly masks activations to prevent co-adaptation~\cite{srivastava2014dropout}; \emph{NoisyNets} injects trainable noise into weight tensors to encourage exploration and stabilize reinforcement learning updates~\cite{fortunato2017noisy}; adversarial training augments data with worst-case perturbations to yield provably robust classifiers~\cite{madry2018towards}; and denoising autoencoders corrupt inputs so networks learn invariances central to self-supervised objectives~\cite{vincent2008extracting}.  Together, these techniques show that training with well-modeled noise yields flatter minima and stronger test-time robustness.

\textbf{From generic perturbations to system-specific noise.}
The unifying principle behind training under controlled noise extends to system constraints. Over the past decade, noise injection has evolved from stochastic regularizers toward modeling \emph{hardware-induced} perturbations. QAT is an outcome of this progression, treating low-precision arithmetic as structured noise injected during every forward pass.
 %This approach not only preserves accuracy, but also aligns optimization with deployment budgets, making robustness a first-class objective for resource-constrained systems.

\subsection{Quantization as a Noise-Injection Process}
Quantization maps a full-precision scalar $x$ to a discrete value
$\tilde{x}=Q_{\!\Delta}(x)=\operatorname{round}(x/\Delta)\,\Delta
       \;=\;x+\epsilon$,
where $\Delta$ is the quantization step (i.e.\ the least-significant bit)
and $\epsilon\!\in[-\Delta/2,\Delta/2]$ is the \emph{quantization error}.
Thus, the forward pass is equivalent to a bounded noisy channel that
smooths the objective to improve robustness.  Contemporary QAT
methods exploit this view by replacing non-differentiable rounding with
differentiable noise proxies—e.g.\ stochastic rounding or uniform noise
whose scale is \emph{learned}—so gradients remain informative while the
network discovers each parameter’s noise tolerance~\cite{fan2020quantnoise,savarese2022not}.
The learned noise scales map directly to mixed bit-width assignments that satisfy strict memory–latency envelopes.

% , and empirical studies from CNNs
% to quantum circuits show that such noise-proxy optimization can deliver
% up to $16\times$ compression and multi-$\times$ speed-ups with negligible
% accuracy loss~\cite{wang2022quantumnat,jacob2018quantization}. \cyrus{I thikn here we only talk about quantum}

\subsection{System Constraints for Quantized Inference}
Low-precision networks (where weights and activations are $\leq$8-bit)  must meet specific system constraints to be suitable for deployment on mainstream systems.  
%We index them \textbf{(C1)}–\textbf{(C5)} for later reference.
\vspace{-0.2em}
\begin{enumerate}[label=\textbf{(C\arabic*)},leftmargin=*]
  \item \textbf{Uniform precision inside a SIMD/tile block.}
        Elements in the \emph{same} vector register or tensor-core tile should share one bit-width; mixing precisions forces serial execution or per-lane control, reducing throughput and raising complexity/cost. \label{C1}
\vspace{-0.2em}
  \item \textbf{Shared scale per fused dot-product.}
        Operands that accumulate into a single output (e.g., a dot product) must share one quantization scale; different scales imply on-the-fly dequantization or higher-precision multiplies, which current platforms handle poorly. \label{C2}
\vspace{-0.2em}
  \item \textbf{Small number of precision levels.}
        Use only a few distinct bit-widths (ideally two or three) so kernels and hardware support a small set of patterns without heavy runtime dispatch or added hardware complexity. \label{C3}
\vspace{-0.4em}
  \item \textbf{Hardware-friendly encoding/decoding.}
        Precision encoding/decoding should feed the MAC datapath directly—no extra masking or looping—otherwise overheads can negate quantization gains. \label{C4}
\vspace{-0.2em}
  \item \textbf{Vector/Tile-friendly memory layout.}
        Lay out quantized tensors so a full SIMD word is fetched with one coalesced load (e.g., channel-packed or row-blocked) to minimize memory traffic. \label{C5}
\end{enumerate}

Our \FWName{} framework integrates system-level constraints into the training process of quantized neural networks and provides corresponding software, hardware support and optimization during inference. This holistic co-design approach enables aggressive quantization while preserving model accuracy and ensures that system constraints are met to maximize inference efficiency.

\section{\FWName{}-QAT: System-Aware Quantization-Aware Training with Structured Noise Injection and Normalization-Aware Noise Scaling} \label{sec:titan-qat}

In this section, we detail \FWName{}-QAT, a quantization-aware training algorithm that surpasses full-precision performance on both convolutional and transformer models (Complete algorithm is outlined in Appendix \ref{app:qat_algo}, Algorithm~\ref{alg:titan_qat}). \FWName{}-QAT is the core of our \FWName{} framework. It is expressly {system/hardware-aware}, where the following hyperparameters are mapped to concrete costs on a given target inference platform. Consequently, models trained with \FWName{}-QAT are more efficient during inference on the target hardware.
    \vspace{-0.2em}
\[
  \left(
    k,\,
    \tau_{\mathrm{final}},\,
    \lambda,\,
    \mathrm{GroupSize}_{\mathrm{act}},\,
    \mathrm{GroupSize}_{\mathrm{wt}}
  \right)
      \vspace{-0.2em}
\]

%each of which controls a distinct hardware bottleneck:
\begin{itemize}
  \item $k$ -- maximum \emph{number of precision levels}.  
        We constrain this number to avoid high overheads associated with encoding/decoding of precision levels, hardware computation units supporting mixed-precision operations, and/or data shuffling through software.
    \vspace{-0.5em}
  \item $\text{GroupSize}_{\text{act}},\ \text{GroupSize}_{\text{wt}}$ --
        sizes of SIMD-aligned blocks whose elements share one
        scale/bit-width; they are set to be positive integer multiples of the vector
        length of the inference engine (e.g.\ 64 for AVX512 FP8 \cite{intel_avx512} or 128 for
        TPU-v4 INT8 \cite{google_edgetpu}).
    \vspace{-0.5em}
  \item $\tau_{\text{final}}$ -- annealing \cite{jang2016categorical} schedule end-point; larger $\tau_{\text{final}}$ values tighten the softmax in Line~4, nudging each \emph{channel}
        toward a single bit-width earlier and reducing costly
        noise-gradient updates.
\vspace{-0.5em}
  \item $\lambda$ -- coefficient of the $\ell_{1}$ regularizer in Line~35
        that discourages high precisions; its value can be adjusted to meet
        the desired efficiency–accuracy trade-off.
\end{itemize}

\begin{figure}[b]
    \centering
    \vspace{1.5em}
    \includegraphics[width=1.0\linewidth]{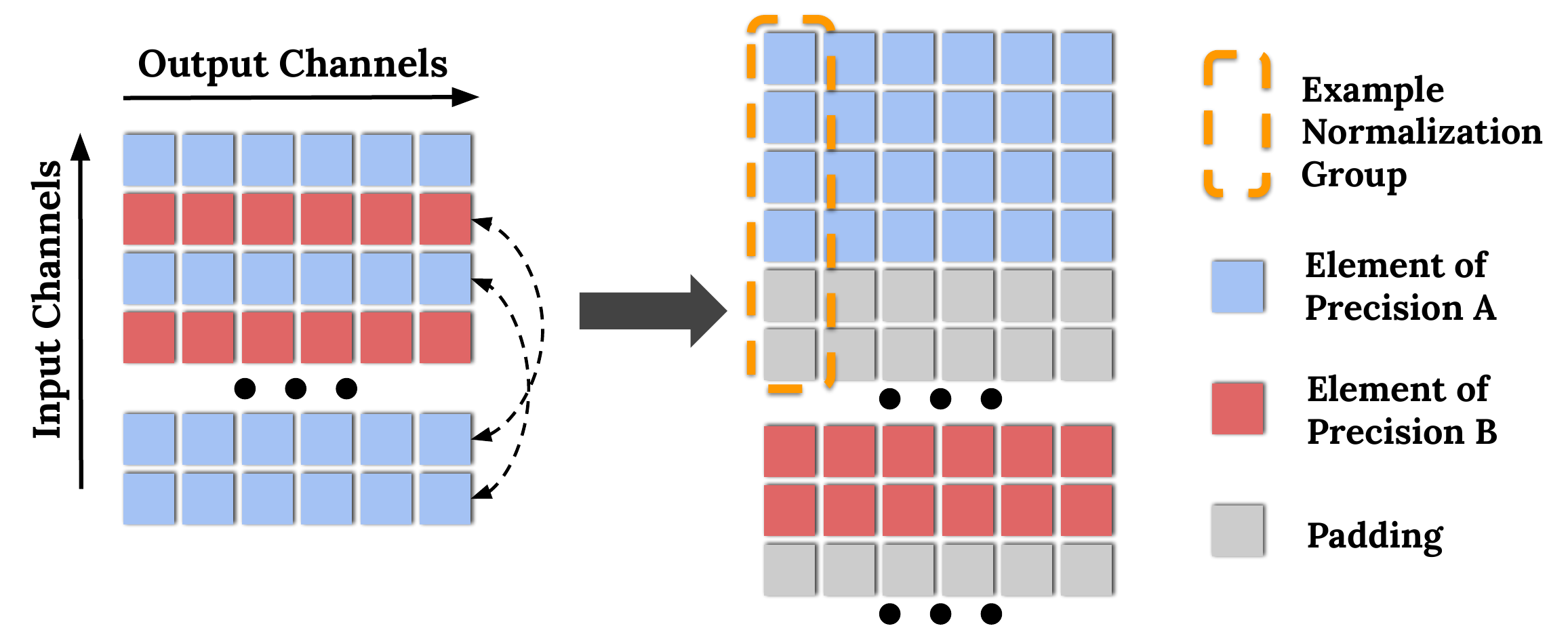}
        \vspace{-1.5em}
\caption{\FWName{}-QAT inference-aware normalization: group channels by bit-width into uniform-precision blocks, then pad each block to the nearest SIMD width to maximize vector utilization without changing results.}

    % \Description{Visualization of normalization process in \FWName{-QAT}, showing grouped channels and padding to SIMD vector length.}
    \label{fig:normalization}
\end{figure}

\vspace{-1em}
\paragraph{Incorporating System Constraints into \FWName{}-QAT.}
Alg.~\ref{alg:titan_qat} embeds the §2.4 constraints (C1–C5) as tunable design knobs, steering training toward accurate yet efficient configurations.

\vspace{-0.5em}
\begin{itemize}[leftmargin=*]
  \item \textbf{C1 — Uniform precision per SIMD/tensor block.}
        Set \(\text{GroupSize}_{\text{wt}}\) and \(\text{GroupSize}_{\text{act}}\) to multiples of the vector width; the permutation in \textbf{Line \ref{ln:permute}} clusters equal bit-width channels. Groupwise noise updates (L11–22) match this granularity.
    \vspace{-0.7em}
  \item \textbf{C2 — Shared scale per fused dot-product.}
        Assign one scale per weight/activation block \((w_{\max}^{(l,g)}, x_{\max}^{(l,g)})\) in L11–22, eliminating run-time dequantization.
    \vspace{-0.7em}
  \item \textbf{C3 — Small number of precision levels.}
        \(k\) (L1–2) upper-bounds available bit-widths; an annealed softmax (L4, used in L7) sharpens choices; final rounding (L28–30) enforces \(\le k\) levels.
    \vspace{-0.7em}
  \item \textbf{C4 — Hardware-friendly encoding/decoding.}
        Fixed-point operand formats (§\ref{sec:codec}) feed MACs directly, so each fused dot-product executes as a pure fixed-point MAC.
    \vspace{-0.7em}
  \item \textbf{C5 — Vector-friendly memory layout.}
        The permutation (Line \ref{ln:permute}) plus padding to \(\text{GroupSize}\) coalesces SIMD loads; the layout is fixed in Phase I (L10, L11–22) and reused at inference.
\end{itemize}

% \vspace{-0.47em}
\subsection*{Phase I: Noise-Driven Bit-Width Discovery}
For every \emph{layer} $l$, learn a \emph{channel-wise} scale vector
% \vspace{-0.3em}
% \vspace{-0.4em}
\[
  s^{(l)}=\operatorname{softmax}\!\left(\tau\,z^{(l)}\right)\,\odot\,v
  \;\in\;\mathbb{R}^{d_l},
\]
then, for each channel $i$, inject noise proportional to
$\sigma\!\left(s^{(l)}_{i}\right)=2^{s^{(l)}_{i}}$:
\[
  w^{(l)}_{i} \leftarrow
  w^{(l)}_{i} +
  \varepsilon^{(t)}_{w_i}\,
  \sigma\!\left(s^{(l)}_{i}\right)\,
  w_{\max}^{(l,g)},\qquad
  \varepsilon^{(t)}_{w_i}\sim\mathcal{U}\{-1,1\},
\]
\vspace{1em}
with a similar rule for activations (Lines 11–22).  
Both weights and activations are processed \emph{group-wise}, where each
group $g$ contains exactly $\text{GroupSize}_{\text{wt}}$ or
$\text{GroupSize}_{\text{act}}$ contiguous elements/channels, so that their scales can be fused into a single SIMD multiply during inference.

\paragraph{Dynamic channel re-ordering.}
After every noise update, channels are permuted so that those with
identical arg-max precision become adjacent.  
This yields an inference-ready layout where channels with the same precision
are stored contiguously, eliminating scatter/gather memory accesses.  
The strategy follows the principle that quantization preserves accuracy
best when normalization is applied to groups of parameters that share similar
perturbation sensitivities as revealed by their learned precision.

\paragraph{Precision-palette regularization.}
The loss
\[
  \mathcal{L}
  \;=\;
  L\!\left(w+\text{noise}\right)\;+\;
  \lambda\left\lVert
    \log_{2}\!\left(1+e^{-s}\right)
  \right\rVert_{1}
\]
combines task error with an $\ell_{1}$ penalty on the \emph{effective
bit-width} $b(s)=\log_{2}(1+e^{-s})$.  The soft cardinality constraint
induced by the $\ell_{1}$ norm caps the maximum number of precision levels to $\le k$. 

% \vspace{0.3em}
\subsection*{Phase II: Quantized Fine-Tuning (Lines 32-37)}
The final bit-width of each channel is
\[
  p^{(l)}_{i}
  \;=\;
  1+\operatorname{round}\!\left(\log_{2}(1+e^{-s^{(l)}_{i}})\right),
  \quad
  p^{(l)}_{i}\le 8,
\]
The 8-bit ceiling aligns with mainstream integer hardware paths. We quantize \(w\) and \(x\) once per batch and train with a straight-through estimator (STE)~\cite{bengio2013estimating}, which backpropagates through the non-differentiable quantizer as identity.

\subsection*{Enabling Low-Cost Fixed-Point Operations}\label{sec:codec}

All tensors are quantized \textbf{per group} so that a single scale factor can be shared by $G_{\text{act}}$ activations or $G_{\text{wt}}$ weights.

\paragraph{1. Group-wise scale selection}  
For every group $g$ we store a scale based on the absolute maximum value in the group:
\vspace{-0em}
\[
  s_g = \max_{x\in g} |x|.
  \vspace{-0.5em}
\]

% This ensures that scaling is based on the actual maximum value in the group, rather than a power-of-two approximation. The scaling factor is then used for quantization.

\paragraph{2. Activation representation using unsigned integers}  
Activations are mapped to the $n$-bit unsigned integer range $[0,2^{n}-1]$:
\vspace{-0.5em}
\[
  b = \operatorname{clip}\left(\operatorname{round}\left(
        \tfrac{a}{s_g}\,(2^{n}-1)
      \right),\,0,\,2^{n}-1\right)\in\{0,1\}^{n}.
\]
%where $n$ is the number of quantization bits ($n\!\le\!8$ in our design to ensure training stability). 
Here \(\operatorname{clip}\) saturates over-/under-flows.  
\emph{Decoding} is the inverse affine transform
\vspace{-0.5em}
\[
  \hat a \;=\; s_g\,\tfrac{b}{2^{n}-1}.
\]

\paragraph{3. Weight representation using signed two’s-complement fraction}  
Weights are normalized by its group scale, \(w' = w/s_g\), so \(w'\in[-1,+1]\).  
We then choose the closest fixed-point representation of the form
\vspace{-0.5em}
\[
  q(\mathbf{b}) \;=\;
  -b_{n-1} + \sum_{i=0}^{n-2} b_i\,2^{-(i+1)}, 
  \quad \mathbf{b}\in\{0,1\}^{n},
\tag{1}\label{eq:twos}
\]
i.e.\ the usual two’s-complement representation whose MSB is the sign ($-1$) and the remaining bits are fractional powers of two.  The optimal representation is
\vspace{-0.2em}
\[
  \mathbf{b}^{\star} = 
  \arg\min_{\mathbf{b}\in\{0,1\}^{n}}
  \left|\,w' - q(\mathbf{b})\right|.
\]

Bits \(\mathbf{b}^{\star}\) are stored following the format in Eq.~\eqref{eq:twos}.  
\emph{Decoding} during inference simply involves re-evaluating Eq.~\eqref{eq:twos} and rescaling:
\[
  \hat w = s_g \, q(\mathbf{b}^{\star}).
\]

%\paragraph{4. Hardware cost}  
Since the scale factor \(s_g\) is based on the absolute maximum value, all scaling operations can be reduced to integer bit-shifts, which are efficient for hardware implementation. The representation of weights via two's complement allows for operations using integer arithmetic, avoiding the need for floating-point units, making the approach suitable for use in vector MAC units in various AI inference hardware.

\section{\FWName{} Inference Support and Optimization}
\FWName{} fuses software optimizations with hardware support to accelerate models trained with \FWName{}-QAT. This holistic design maximizes end-to-end inference performance in terms of both latency and power efficiency. For the evaluation of this paper, we focus on CPUs and GPUs, though our technique also generalizes to AI hardware in general.

\subsection{Software Optimizations}

\subsubsection{GPU optimizations.}
We employ the hardware-aware scheduler of \citet{zhang2023hardware}.
Given a GPU description and the network’s computational graph, the
scheduler
(i) partitions the graph into tasks that fit within the register file
and on-chip buffers of a single SM, then
(ii) performs a heuristic search over tile sizes, loop nest orderings,
dataflows, and double-buffering strategies to minimize end-to-end
latency.

\subsubsection{CPU–SIMD optimizations.}
For CPUs, we generate custom SIMD kernels with the framework of
\cite{zhou2023simd}, applying 3 key techniques:

(1) \textbf{NCKHWc layout}—channels are blocked by the
      SIMD vector width ($c$), maximizing vector utilization
      \cite{liu2019optimizing,chen2018tvm,zhou2023simd};

(2) \textbf{Kernel fusion}—\textit{Conv2D}, \textit{BatchNorm}, and
      \textit{Quantize} are merged into a single kernel to reduce
      memory traffic \cite{chen2018tvm,zhou2023simd};

(3) \textbf{Dataflow selection}—output-anchored stationarity with
      input- and weight-auxiliary stationarities for $3{\times}3$ convolutions, and
      weight-stationary for $1{\times}1$, the fastest combination on
      modern CPU SIMD back-ends \cite{zhou2023simd}.

% \vspace{0.4em}
\noindent
\textbf{Channel re-ordering and padding.}
For both CPU and GPU back-ends we reorder channels and pad tensors so
that values sharing the same learned precision are stored contiguously,
exactly as in Fig.~\ref{fig:normalization}.

\subsection{Hardware Support}\label{sec:hardware_support}
Commodity CPUs (Intel/AMD AVX-512 VNNI, AMX-INT8) and Arm v8.4-A \texttt{UDOT}/\texttt{SDOT} already perform mixed-sign \(\mathrm{UINT8}\!\times\!\mathrm{INT8}\!\to\!\mathrm{INT32}\) dot products; so do NVIDIA Tensor Cores (8/4-bit) and Google Edge TPU~\cite{intel_vnni,amd_zen4_vnni,arm_usdot,nvidia_tc_int8,nvidia_tc_int4,google_edgetpu}. These ISAs natively match \FWName{}’s data formats. For 1-bit channels, \FWName{} reduces dot products to XNOR+POPCOUNT as in BinaryConnect/BinaryNet/XNOR-Net~\cite{courbariaux2015binaryconnect,courbariaux2016binarynet,rastegari2016xnor}; existing XNOR/POPCOUNT units in CPUs, GPUs, FPGAs (FINN) and accelerators (BitFusion)~\cite{umuroglu2017finn,sharma2018bitfusion} thus support both 1-bit and higher-precision ops in one low-precision pipeline.

\FWName{} therefore requires no new architecture—only a \emph{configurable MAC unit} for low, mixed-precision arithmetic. Many AI MACs already support mixed precision~\cite{nvidia_turing_2018, nvidia_ampere_2020, nvidia_hopper_2022,jouppi_tpu_isca17, jouppi_supercomputer_cacm20, jouppi_tpu_v4_2023, amd_cdna2_2021, apple_coreml_wwdc24, intel_gaudi2_2023, bitfusion_isca18}; for completeness, Appendix~\ref{app:arch} details a design that packs \FWName{} operands into 16-bit lanes and enables configurable 1–8-bit MACs with minimal area/power overhead.

This unit is a drop-in replacement (given ISA support) for: (i) GPU tensor cores, (ii) GPU SIMD lanes, (iii) CPU vector units, and (iv) spatial accelerators (e.g., TPUs, Eyeriss), enabling broad deployment.

Although \FWName{} permits non-power-of-two precisions, §\ref{sec:palette-study} shows two levels suffice when one is \(>\)4-bit; thus supporting 1, 4, and 8 bits via our codec closely matches the empirical optimum across evaluated workloads.

\section{Accuracy-Compression Evaluation Setup}\label{sec:training_setup}

This section details our experimental setup for quantifying the accuracy–compression gains delivered by \FWName{}-QAT. Our study follows the \emph{parsimony principle}: we adopt the coarsest configuration that still reveals the robustness of \FWName{}, thereby minimizing training effort.

\vspace{0.5em}
\noindent\textbf{Normalization and noise scaling.}
\begin{itemize}[leftmargin=*]
\item \textbf{Transformers.} We apply \emph{layer-wise} normalization/noise scaling to the weights, and \emph{group-wise} normalization/noise scaling to the activations using 512-bit groups. Prior work~\cite{xiao2023smoothquant,zhao2023atom} shows that the wide dynamic range of transformer activations benefits from this group-wise treatment.
\item \textbf{CNNs.} Layer-wise normalization and noise scaling are sufficient for both weights and activations.
\end{itemize}

\textbf{Fine-tuning schedule.}
Phase II of Algorithm \ref{alg:titan_qat} follows the DoReFa recipe~\cite{zhou2016dorefa} to enable apples-to-apples comparison with prior work; alternative schedules are drop-in replacements.

\textbf{Benchmarks.} We evaluate \textit{\textbf{(i)}} a 6-layer Transformer encoder–decoder trained on the IWSLT14 German-to-English (De\textrightarrow{}En) dataset~\cite{vaswani2017attention,cettolo2014report}, and \textit{\textbf{(ii)}} three widely used CNNs--PreResNet-18~\cite{he2016deep}, MobileNet-V2~\cite{sandler2018mobilenetv2}, and DenseNet-121~\cite{huang2017densely}--on CIFAR-10~\cite{krizhevsky2009learning}, CIFAR-100~\cite{krizhevsky2009learning}, and ImageNet~\cite{deng2009imagenet}.

\textbf{Training cost.}
For CNNs we run 350 Phase I (noise-injection) epochs followed by 300 Phase II (fine-tuning) epochs, mirroring the unstructured noise-injection baseline~\cite{savarese2022not}. For transformers, we use 36 Phase I and 24 Phase II epochs; the latter can be shortened further with negligible accuracy loss. Fine-grained normalization adds only $\sim$30\% GPU-hour overhead versus~\cite{savarese2022not}, whose training regimen is impractical for real-world deployment, while still being significantly less costly than mixed-precision alternatives such as~\cite{yang2021bsq}. We will open source all experimental hyper-parameters.

% \textbf{Training configurations.} We will open source all hyper-parameters required to reproduce our experimental results. 

\section{Accuracy-Compression Evaluation Results} \label{sec:training_eval}
This section reports the accuracy–compression results of \FWName{} on four benchmarks, demonstrating that it achieves 7–16× model size reductions while matching or surpassing full-precision accuracy. 
\vspace{-0.5em}
\subsubsection*{Table Notations}
In Tables~\ref{tab:transformer_accuracy}–\ref{tab:cifar100_results}, the \textbf{Sys. Fr.} (System-Friendly) column is ticked when a method satisfies all
criteria \textbf{C1}–\textbf{C5} (\S\ref{sec:background}). Footnotes indicate results from implementations; rows without footnotes vary only by the $\ell_{1}$ regularizer $\lambda$.

\vspace{-0.5em}
\subsubsection*{Cross-Benchmark Takeaways}
Compared to SOTA quantization, only \FWName{} (1) matches or exceeds full-precision accuracy, (2) achieves 7–16× compression via joint weight–activation quantization, and (3) maintains a single uniform fixed-point per vector for easy deployment on existing hardware.

%These quantized models run faster, use less memory, and still beat state-of-the-art INT8 or floating-point baselines.

\vspace{-0.5em}
\subsubsection*{Benchmark-Specific Results}

\newcolumntype{L}[1]{>{\raggedright\arraybackslash}p{#1}}
\newcolumntype{Y}{>{\centering\arraybackslash}X}

\begin{table}[htbp]
    \centering
    \small
\caption{Accuracy–compression results on the IWSLT’14 De$\to$En Transformer: BLEU scores versus average bit-widths for weights and activations. \FWName{} outperforms both the full-precision model and all low-precision baselines, achieving higher BLEU scores while operating at markedly lower precisions. This demonstrates the effectiveness of \FWName{} in maintaining translation quality while reducing computational requirements.}

    \label{tab:transformer_accuracy}
    % \small
    \setlength{\tabcolsep}{4pt}
    \renewcommand{\arraystretch}{1.2}

\begin{tabularx}{\columnwidth}{@{} L{0.4\columnwidth} Y Y Y c @{}}
        \toprule
        \textbf{Method} & \textbf{Weight Prec.\,$\downarrow$} & \textbf{Act. Prec.\,$\downarrow$} & \textbf{BLEU Score}\,$\uparrow$ & \textbf{Sys. Fr.} \\
        \midrule
        \multicolumn{5}{@{}l@{}}{\textbf{IWSLT'14 German to English}} \\
        \midrule
        FP32 & 32\footnotemark[1] & 32\footnotemark[1] & 34.9 & \tick \\
        % \textit{FP16} & \textit{16\footnotemark[1]} & \textit{16\footnotemark[1]} & \textit{???} & \textit{\YesGreen} \\
        SMOL \cite{savarese2022not} & 3.9 & 32\footnotemark[1] & 34.7 &  \\
        SMOL \cite{savarese2022not} & 5.8 & 32\footnotemark[1] & 34.9 & \\
        Lee et al. \cite{lee2022toward} & 8\footnotemark[1] & 8\footnotemark[1] & 34.5 & \tick \\
        % Ours\footnotemark[2] & 4.9 & 4.9 &  34.7 & \YesGreen,  \\
        % Ours\footnotemark[2] & 5.8 & 5.8 & 34.8 & \YesGreen, FP \\
        \textit{Ours\footnotemark[2]} & \textit{5.2} & \textit{5.2} & \textit{34.9} & \tick \\
        \textit{Ours\footnotemark[3]} & \textit{4.9} & \textit{4.9} & \textit{35.2} & \tick  \\
        \textit{Ours\footnotemark[3]} & \textit{6.5} &\textit{ 6.5} & \textit{35.4} & \tick \\
        \bottomrule
    \end{tabularx}
      \vspace{-0.5em}
\end{table}

\paragraph{Transformers.}

\FWName{} attains the highest BLEU score while using the lowest activation precision. With an average of 5.2 bits for both weights and activations, BLEU matches the full-precision baseline; omitting quantization of the final dense layer yields a significant BLEU gain over FP32 implementations at 4.9/6.5-bit weight/activation precision. Table~\ref{tab:transformer_accuracy} shows that \FWName{} surpasses state-of-the-art methods that keep activations in full precision \cite{savarese2022not}, yielding a 54 \% higher compression ratio and a significant 0.9 BLEU advantage over recent floating-point quantizers (\citet{lee2022toward}).

\begin{table}[t]
  \footnotesize
  \centering
  \caption{Accuracy–compression results for ResNet-18 on ImageNet
           (top-1 accuracy vs.\ average bit-width). 
           \FWName{} leads at low precision, surpassing both full- and low-precision baselines.}
  \label{tab:imagenet_results}
  \setlength{\tabcolsep}{4pt}
  \renewcommand{\arraystretch}{1.15}

  \begin{tabularx}{\columnwidth}{@{} L{0.4\columnwidth} Y Y Y c @{}}
    \toprule
    \textbf{Method} &
    \textbf{Weight Prec.\,$\downarrow$} &
    \textbf{Act.\ Prec.\,$\downarrow$} &
    \textbf{Acc.\,$\uparrow$} &
    \textbf{Sys.\,Fr.} \\
    \midrule
    FP32 & 32\textsuperscript{1} & 32\textsuperscript{1} & 69.6 & \tick \\
    SMOL \cite{savarese2022not} & 4.20 & 32\textsuperscript{1} & 70.4 & \\
    SMOL \cite{savarese2022not} & 4.50 & 32\textsuperscript{1} & 70.6 & \\
    PACT \cite{choi2018pact} & 32\textsuperscript{1} & 4 & 69.2 & \\
    DSQ \cite{gong2019differentiable} & 4 & 32\textsuperscript{1} & 69.6 & \\
    Bit-Split \cite{wang2022optimization} & 8 & 8 & 69.7 & \tick \\
    \addlinespace[0.3em]
    \textit{Ours} & \textit{6.03\textsuperscript{4}} & \textit{6.03\textsuperscript{4}} & \textit{70.2} & \tick \\
    \textit{Ours} & \textit{6.25} & \textit{6.25} & \textit{70.8} & \tick \\
    \bottomrule
  \end{tabularx}
  % \vspace{-1.5em}
\end{table}

\vspace{-1em}

\paragraph{ImageNet (ResNet-18).}
Table \ref{tab:imagenet_results} shows that \FWName{} tops the top-1 accuracy ranking with both quantized weights and activations,  while all other prior methods except for Bit-Split \cite{wang2022optimization}, with the nearest rival being SMOL \cite{savarese2022not}, cannot quantize both weights and activations but still yield lower accuracy than \FWName{}. \FWName{} outperforms Bit-Split in compression ratios of both weights and activations, and top-1 accuracy.

\newcolumntype{Y}{>{\centering\arraybackslash}X}
\newcolumntype{L}[1]{>{\raggedright\arraybackslash}p{#1}}

% \paragraph{CIFAR-10.}

% On \textbf{PreResNet-18}, \FWName{} achieves the best accuracy–size trade-off: only SMOL slightly surpasses its accuracy, but it keeps activations in full precision, whereas \FWName{} compresses them to $\le 3.17$ bits on average. Retaining depth-wise convolutions in high precision allows \FWName{} to reach $2.2$-bit weights and activations—yielding a $14.5\times$ model-size reduction with no accuracy loss. Because depth-wise layers are the main bottleneck, we apply group-wise normalization or preserve higher precision for those filters.

% \noindent On \textbf{MobileNet-v2} and \textbf{DenseNet-121}, \FWName{} is the only method that jointly quantizes weights \emph{and} activations while matching or surpassing every competitor and the FP32 baseline. To the best of our knowledge, this is the first reported joint-quantization result for DenseNet-121, achieving full-precision accuracy with just $3.97$-bit weights/activations on average.
\paragraph{CIFAR-10.}

\textbf{(1) PreResNet-18.} \FWName{} sets a new Pareto front: with a 2.23-bit palette it attains \textbf{95.3\%} accuracy (\(+0.2\) pp vs.\ FP32) at a \(32/2.23 \!\approx\! 14\times\) size reduction.  Pushing to 1.98 bits still preserves 95.0\%.  Keeping depth-wise convolutions in higher precision (or applying group-wise normalization) is essential; removing this safeguard drops accuracy by \(>\!1\) pp, confirming them as the primary quantization bottleneck. \textbf{(2) MobileNet-v2.}  At 2.20 bits, \FWName{} exactly matches FP32 accuracy (94.2\%) while compressing the model by \(\approx14.5\times\).  A 3.17-bit variant yields 93.8\%, far above Liu et al.\ (84.8\% at 3.3 bits).  Competing methods like SMOL and BSQ leave activations in FP32, forfeiting both memory and latency gains.
\textbf{(3) DenseNet-121.}  To our knowledge, our results are the first joint-quantization numbers for this architecture.  \FWName{} preserves full-precision accuracy (94.3\%) at 3.97 bits smaller); even an aggressive 1.86-bit setting still delivers 93.6\% with a \(\approx17\times\) reduction, requiring no model-specific tuning.

% Across all three networks, \FWName{} is the only approach that drives \emph{both} weights \emph{and} activations below 4 bits, satisfies all system-compatibility constraints, and matches or exceeds FP32 accuracy.
\vspace{-1.5em}

\begin{table}[htbp]
  \footnotesize\centering
  \caption{Accuracy–compression results on CIFAR-10 for three architectures.
           \FWName{} matches full-precision accuracy at considerably lower
           precision while remaining system-friendly.}

    \vspace{1em}
  \label{tab:cifar10}
  \setlength{\tabcolsep}{4pt}          % reduce if still tight
  \renewcommand{\arraystretch}{1.15}

  \begin{tabularx}{\columnwidth}{@{} p{1.1in} *{4}{>{\centering\arraybackslash}X} @{}}
    \toprule
    \textbf{Method} & \textbf{Weight Prec.\,$\downarrow$} &
    \textbf{Act.\ Prec.\,$\downarrow$} & \textbf{Acc.\,$\uparrow$} &
    \textbf{Sys.\,Fr.} \\ \midrule

    \multicolumn{5}{@{}l@{}}{\textbf{MobileNet-v2}} \\ \midrule
    FP32 & 32\textsuperscript{1} & 32\textsuperscript{1} & 94.2 & \tick \\
    SMOL \cite{savarese2022not}  & 1.5 & 32\textsuperscript{1} & 94.5 & \\
    SMOL \cite{savarese2022not}  & 1.7 & 32\textsuperscript{1} & 94.8 & \\
    Liu et al.\ \cite{liu2021layer} & 3.32 & 3.39 & 84.8 & \\
    BSQ \cite{yang2021bsq}       & 2.8 & 32\textsuperscript{1} & 94.1 & \\
    \addlinespace[0.3em]
    \textit{Ours}\textsuperscript{5} & \textit{2.20} & \textit{2.20} & \textit{94.2} & \tick \\
    \textit{Ours} & \textit{3.17} & \textit{3.17} & \textit{93.8} & \tick \\

    \midrule
    \multicolumn{5}{@{}l@{}}{\textbf{Preact ResNet-18}} \\ \midrule
    FP32 & 32\textsuperscript{1} & 32\textsuperscript{1} & 95.1 & \tick \\
    Yin et al.\ \cite{yin2018binaryrelax} & 2 & 32\textsuperscript{1} & 95.0 & \\
    Liu et al.\ \cite{liu2021layer}      & 3.07 & 4.35 & 94.1 & \\
    \addlinespace[0.3em]
    \textit{Ours} & \textit{2.23} & \textit{2.23} & \textit{95.3} & \tick \\
    \textit{Ours} & \textit{1.98} & \textit{1.98} & \textit{95.0} & \tick \\

    \midrule
    \multicolumn{5}{@{}l@{}}{\textbf{DenseNet-121}} \\ \midrule
    FP32 & 32\textsuperscript{1} & 32\textsuperscript{1} & 94.3 & \tick \\
    \textit{Ours} & \textit{3.97} & \textit{3.97} & \textit{94.3} & \tick \\
    \textit{Ours} & \textit{1.86} & \textit{1.86} & \textit{93.6} & \tick \\

    \bottomrule
  \end{tabularx}
    \vspace{-3em}
\end{table}

\paragraph{CIFAR-100.}
As reported in Table~\ref{tab:cifar100_results}, \FWName{} on PreResNet-18 matches the full-precision baseline at a 16× compression ratio.  On MobileNet-v2, \FWName{} surpasses the FP32 baseline and—while trailing LSQ~\cite{esser2019learned} by just 0.2 pp—uses a 2.89-bit palette versus LSQ’s 4 bits, delivering a 38.4\% higher compression ratio.   It also beats the 2-bit LQW+SBR and LQW+CAQ baselines by 3.7 pp and 6.7 pp, respectively, remaining the only sub-3-bit method that satisfies all system criteria.

% \vspace{-0.5em}

% \vspace{-1em}

% \vspace{-1.5em}
\begin{table}[htbp]
    \centering
    \small
    \caption{Accuracy–compression trade-offs on CIFAR-100: test accuracy versus average bit-width. \FWName{} matches full-precision accuracy at 2-bit precision on PreResNet-18 and outperforms all low-bit baselines on MobileNet-v2 while using fewer bits.}
      \vspace{1em}

    \label{tab:cifar100_results}
    \setlength{\tabcolsep}{5pt}
    \renewcommand{\arraystretch}{1.2}
    \begin{tabularx}{\columnwidth}{@{} p{1.1in} *{4}{>{\centering\arraybackslash}X} @{}}
        \toprule
        \textbf{Method} & \textbf{Weight Prec.\,$\downarrow$} & \textbf{Act. Prec.\,$\downarrow$} & \textbf{Accuracy $\uparrow$} & \textbf{System-Friendly} \\
        \midrule
        \multicolumn{5}{l}{\textbf{MobileNet-v2}} \\
        \midrule
        FP32 & 32\footnotemark[1] & 32\footnotemark[1] & 72.1 & \tick \\
        LSQ \cite{esser2019learned} & 4 & 4 & 72.4 &  \tick \\
        LQW+SBR \cite{hoang2020direct} & 2 & 2 & 68.5 & \tick \\
        LQW+CAQ \cite{hoang2020direct} & 2 & 2 & 65.5 & \tick \\
        \addlinespace[0.35em] 
        \textit{Ours} & \textit{2.89} & \textit{2.89} & \textit{72.2} & \tick \\ 
        \midrule
        \multicolumn{5}{l}{\textbf{Preact ResNet-18}} \\
        \midrule
        FP32 & 32\footnotemark[1] & 32\footnotemark[1] & 75.8 & \tick \\
        \addlinespace[0.35em] 
        \textit{Ours} & \textit{2.00 }& \textit{2.00} & \textit{75.8} & \tick \\
        \bottomrule
    \end{tabularx}
    \footnotetext[1]{Floating-point precision.}
    \vspace{0.2em}
\end{table}

\footnotetext[1]{Parameters in floating point representation.}
% \footnotetext[2]{Final output projection layer in FP16.}
\footnotetext[2]{Final dense layer's weights in \FWName{} format and act. in FP32.}
\footnotetext[3]{Final dense layer's weights and activations both in FP32.}
\footnotetext[4]{Did not constrain both precisions to be $\leq8$.}
\footnotetext[5]{Depthwise Convolutions kept in FP32.}

\begin{figure}[!t]
\centering
    \begin{subfigure}[t]{0.49\columnwidth}
        \centering
        \includegraphics[width=\linewidth]{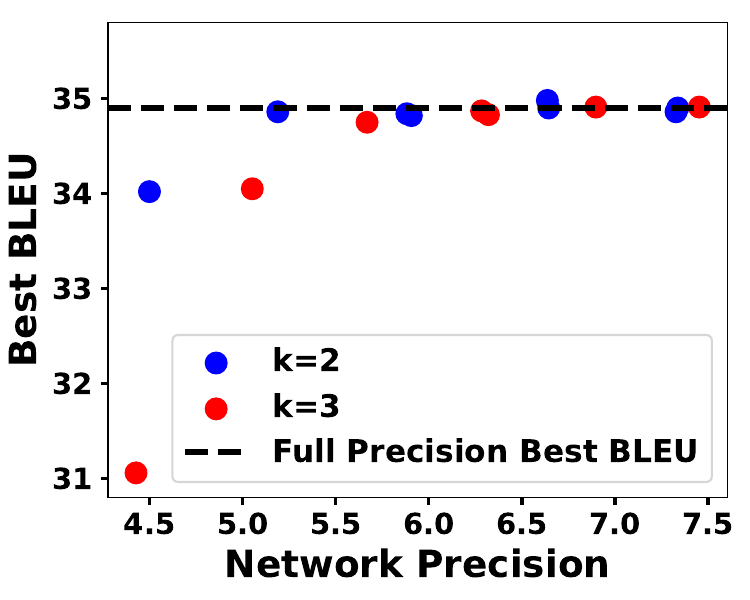}
        \vspace{-1em}
         \footnotesize{\caption{Transformer, IWSLT14'. Last layer in \FWName{} formats.}\label{fig:transformer_k2_vs_k3_allquant}}
        
    \end{subfigure}
    \hfill    
    \begin{subfigure}[t]{0.49\columnwidth}
        \centering
        \includegraphics[width=\linewidth]{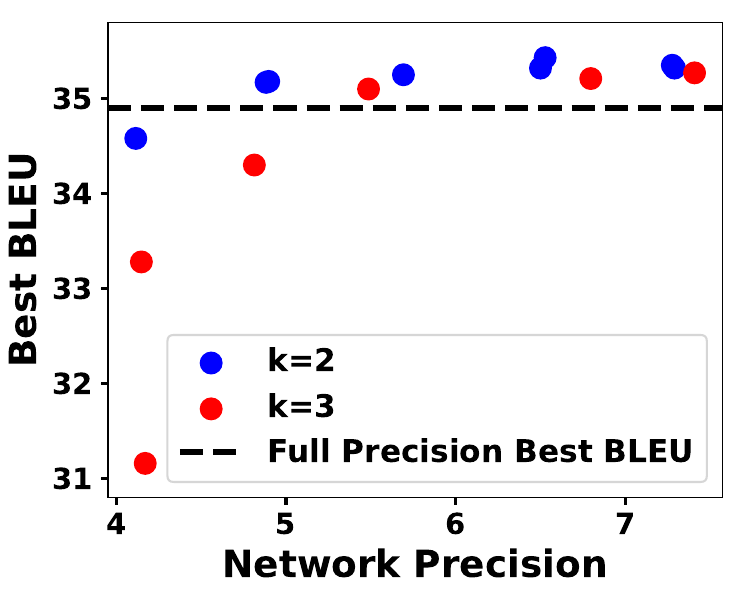}
        \vspace{-1em}
        \footnotesize{\caption{Transformer, IWSLT14'. Last layer weights in FP32.}\label{fig:transformer_k2_vs_k3_nolast}}
        
    \end{subfigure}
    \vskip\baselineskip
    \vspace{-1em}
    % First Row
    \begin{subfigure}[t]{0.49\columnwidth}
        \centering
        \includegraphics[width=\linewidth]{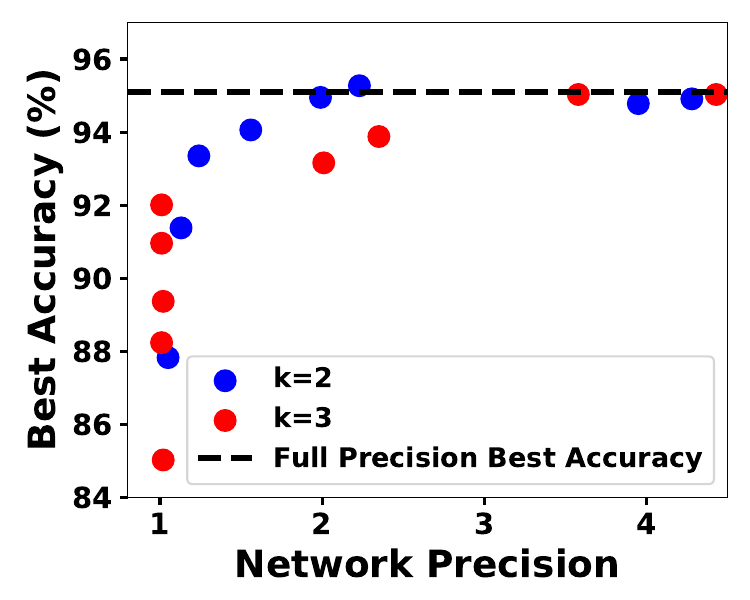}
        \vspace{-1em}
        \footnotesize{\caption{PreResNet-18, CIFAR-10.} \label{fig:preresnet_k2_vs_k3_cifar10}}
       
    \end{subfigure}
    \hfill
    \begin{subfigure}[t]{0.49\columnwidth}
        \centering
        \includegraphics[width=\linewidth]{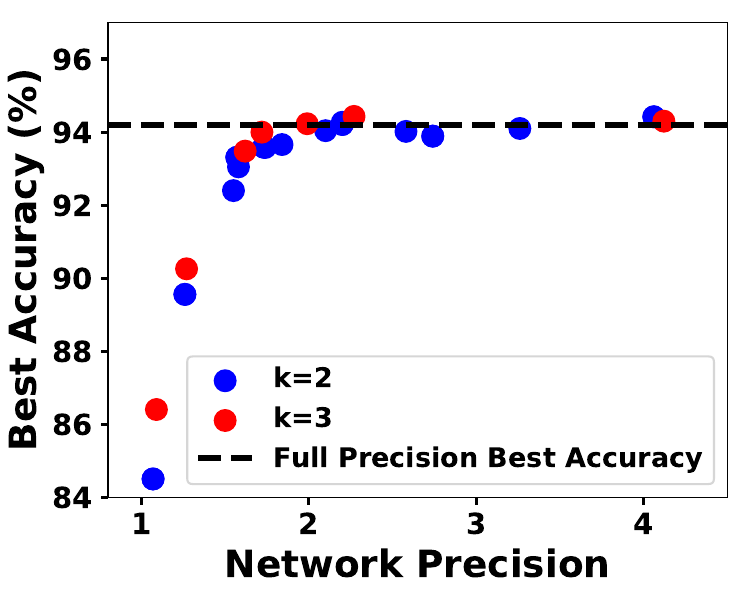}
        \vspace{-1em}
        \footnotesize{\caption{MobileNet-v2\footnotemark[5], CIFAR-10.}\label{fig:figure2}}
        
    \end{subfigure}
    \vskip\baselineskip
    \vspace{-1em}
    % Second Row
    % \begin{subfigure}[t]{0.48\columnwidth}
    %     \centering
    %     \includegraphics[width=\linewidth]{figures/densenet_k2_vs_k3_cifar10.pdf}
    %     \caption{\footnotesize{DenseNet-121 on CIFAR-10}}
    %     \label{fig:figure3}
    % \end{subfigure}    

        \begin{subfigure}[t]{0.49\columnwidth}
        \centering
        \includegraphics[width=\linewidth]{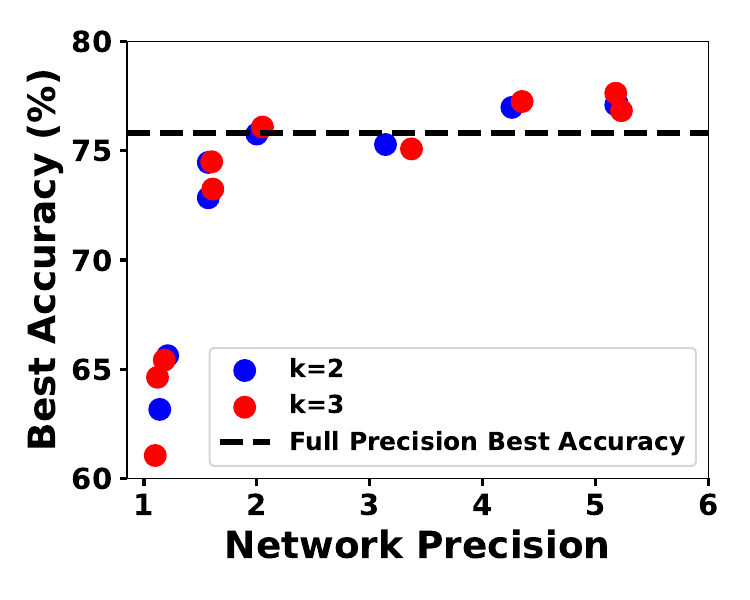}
        \vspace{-1em}
        \footnotesize{\caption{PreResNet-18, CIFAR-100.}\label{fig:figure5}}
        
    \end{subfigure}
    \hfill
    \begin{subfigure}[t]{0.49\columnwidth}
        \centering
        \includegraphics[width=\linewidth]{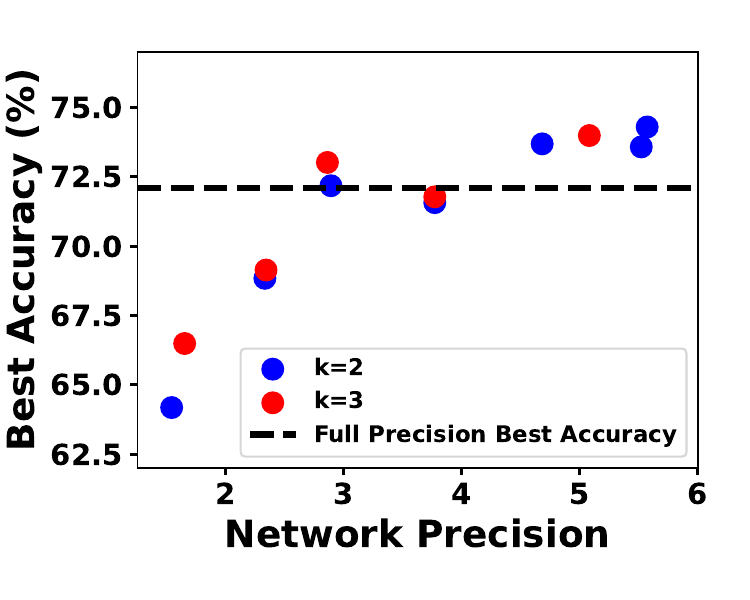}
        \vspace{-1em}
        \footnotesize{\caption{MobileNet-v2, CIFAR-100.}\label{fig:transformer_placeholder}}
        
    \end{subfigure}

    \vspace{0.1em}
    \caption{Average weight–activation precision vs. accuracy for networks with k = 2 and k = 3 precision levels; k = 2 already achieves the optimal accuracy–compression tradeoff.}
    % \vspace{-2.2em}
    \label{fig:k2vsk3}

\end{figure}

\section{Studies on Quantization Precisions} \label{sec:palette-study}

We empirically examine and analyze the \emph{precision palette}, defined as the set of bit-widths a model may select during training using \FWName{}-QAT.  Based on these results, we reveal three new key findings regarding the properties of optimized quantization of neural networks.  %: \textit{(i)} how many levels suffice, \textit{(ii)} which levels matter most, and \textit{(iii)} whether a fixed palette harms accuracy. The subsections below summarize the resulting insights.

\paragraph{Finding I: Two precision levels are generally sufficient.}
We analyzed the impacts of varying the hyper-parameter $k$, which is used to constrain the maximum number of precision levels in \FWName{}-QAT. 
We ran experiments using (\(k\in\{2,3\}\)) per model and plotted accuracy versus average bits-per-parameter (Fig.~\ref{fig:k2vsk3}).  Across all CNN and transformer runs, \(k=2\) matched—or occasionally exceeded—the accuracy of \(k=3\) under the same training budget (see also Figs.~\ref{fig:transformer_k2_vs_k3_allquant}–\ref{fig:preresnet_k2_vs_k3_cifar10}).  When setting \(k=3\), optimization invariably collapsed to either (i) two distinct precisions or (ii) three precisions, but two of which are very close in value,  which indicates vanishing returns beyond \(k=2\).\label{subsec:palette-size}

\paragraph{Finding II: The winning palette is consistently bimodal--one high, one low—across various benchmarks.}
Without any hand-crafted bias, networks trained with \FWName{}-QAT gravitated to a coarse–fine pair of bit-widths.  
\emph{Transformers} consistently settled on \(\{4,8\}\)-bit, while \emph{CNNs} on CIFAR-10/100 adopted \(\{1,\;5\text{–}8\}\)-bit mixes.  
We hypothesize that this clear pattern arises because a wider gap enables the optimizer to allocate higher precision to quantization-sensitive channels, while driving the remaining values toward extremely low precision—more effectively minimizing the loss function that considers both accuracy and bit-widths.  This pattern further suggests that AI hardware may only require two well-optimized datapaths, one sub-byte and one byte-aligned for the two precision levels respectively, to efficiently support the majority of practical workloads.\label{subsec:palette-composition}

\vspace{-1em}

\begin{figure}[htbp]
    \centering
    \includegraphics[width=0.85\linewidth]{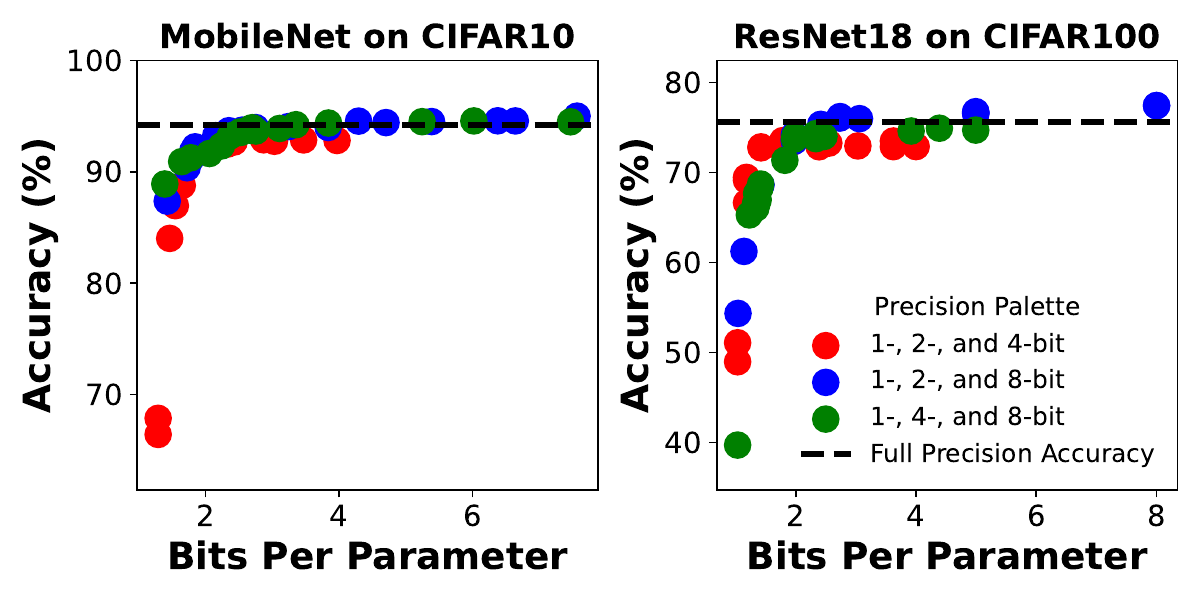}
    \vspace{-1em}
\caption{Accuracy vs. bits/param for \FWName{} with fixed palettes. A single $>$4-bit option preserves full-precision accuracy -- power-of-two palettes incur negligible loss.}
    \label{fig:fixed_palette}
        % \vspace{-1em}
\end{figure}

\paragraph{Finding III: Constraining the precision palette is benign if it includes a high-precision option.}
In practice, one may restrict to a fixed set (e.g., powers of two). We evaluate three palettes: (i) 1/2/4-bit, (ii) 1/2/8-bit, and (iii) 1/4/8-bit. As Figure \ref{fig:fixed_palette} shows, \FWName{} still matches full-precision accuracy at very low average bits per parameter whenever the palette includes a $>$4-bit level—yielding accuracy on par with the flexible, learned-precision strategy. \label{subsec:palette-fixed}

 % \{1,2,4\}, \{1,2,8\}, and \{1,4,8\}

% \section{Software Support}

\vspace{-0.5em}

\section{Inference Evaluation Setup} \label{sec:inference_setup}
%The precision distributions of the trained \FGMPNN{}s and hardware information are provided to a code generator. The code generator leverages such information and combines it with various optimization techniques to output efficient inference kernels.

Practical deployment of neural networks requires careful inference optimizations, and the optimization configurations can be largely different depending on the workload and the underlying hardware platform. We perform proper inference optimizations for both \FGMPNN{s} and baselines in our simulations to genuinely reflect the practical deployment.  These techniques not only enhance performance but also provide insights into how \FWName{} can be effectively utilized and adapted for different hardware environments.

\subsection{Evaluation Methodology}
\subsubsection{GPU Evaluation}
We estimate latency with an analytical model derived from LLMCompass~\cite{zhang2023hardware} (10.4\,\% operator, 4.1\,\% end-to-end error), extended with GEMM for vector units, MAC-aligned compute modules, and a cost model where weight-only quantization is memory-saving but compute-neutral, yielding accurate projections for \FWName{}. Evaluations use $4\times$ NVIDIA A100 and $4\times$ AMD MI210 on both tensor cores and vector units.

\vspace{-0.5em}

\subsubsection{CPU evaluation.}
For every network-precision pair, we run the optimized kernels
from §\ref{sec:inference_setup} in the cycle-accurate \emph{GEM5} simulator%
~\cite{binkert2011gem5}, and simulation parameters are given in
Table~\ref{tab:gem5}.  Uniform INT8 versions of each workload provide a conservative lower bound on the latency
improvements attainable with \FWName{}, since INT8 already outperforms
FP32 and prior quantizers such as SMOL \cite{savarese2022not}.

\vspace{-0.5em}

We compare \FWName{} with full-precision and prior quantizers to illustrate
the latency–precision–accuracy trade-off.  For transformers, we report
both per-operator and end-to-end GPU latencies, because
non-\texttt{MatMul}/\texttt{Linear} kernels contribute to a substantial
fraction of runtime.  For CNNs, we obtain the latency for convolution layers only, as the delays of other layers are negligible.

\vspace{-0.5em}
Note that, we omit GPU timings for CNNs, as LLMCompass is tailored for transformer models, and exclude CPU timings for transformers due to the prohibitive cost of cycle-accurate GEMS simulations on large models. Nonetheless, the qualitative latency trends hold across various hardware platforms, including dedicated AI accelerators.

\begin{table}[h]
\vspace{-1em}
\small% Reduce font size within the table
\setlength{\tabcolsep}{3pt} % Adjust column separation

\caption{GEM5 Simulation Parameters}
\vspace{0.5em}
\label{tab:gem5}
\centering
\begin{tabularx}{\columnwidth}{>{\centering\arraybackslash}m{2.5cm} >{\RaggedRight\arraybackslash}X}
\toprule
\textbf{Component} & \textbf{Description}  \\
\midrule
CPU & Modified O3 CPU model to match the architecture in Fig.~\ref{fig:cpuandgpusimd}.  \\
L1 I-cache & 16\,KB, 4-way associative \\
L1 D-cache & 64\,KB, 4-way associative \\
L2 Cache & 256\,KB, 8-way associative \\
SIMD Register File & 32 128-bit registers \\
\bottomrule
\end{tabularx}

\end{table}
\section{Inference Evaluation}\label{sec:inference-eval}

\noindent To demonstrate its effectiveness, we evaluate \FWName{} on diverse workloads and report its accuracy, compression ratios, and inference latency.

\medskip
\noindent\textbf{CPU results.} As shown in Table~\ref{tab:cnn_latency_cpu}, the compression ratios achieved by \FWName{} translate into pronounced speedups for CNNs. Across all benchmark networks, \FWName{} delivers up to a \textbf{7.30$\times$} speedup over INT8 baselines that quantize both weights and activations. Even for PreResNet-18 on ImageNet, where the average parameter precision is 6.25\,bits, we still observe a \textbf{2.87$\times$} speedup. This is because the new MAC instructions require fewer cycles to execute and eliminate cross-lane reductions.

% \vspace{-0.5em}
\begin{table}[htbp]
  \centering
  % shrink or stretch so the width equals \columnwidth
  \resizebox{\columnwidth}{!}{%
    \begin{tabular}{l l c c c c}
      \toprule
      \textbf{Architecture} & \textbf{Dataset} & \textbf{Bpp.}$\downarrow$ &
      \textbf{Acc.}$\uparrow$ & \textbf{FP Acc.} & \textbf{Speedup}$\uparrow$ \\
      \midrule
      PreResNet-18 & ImageNet   & 6.25 & 70.8 & 69.6 & 2.87$\times$ \\
      PreResNet-18 & CIFAR-10   & 2.23 & 95.3 & 95.1 & 4.87$\times$ \\
      PreResNet-18 & CIFAR-10   & 1.98 & 95.0 & 95.1 & 7.15$\times$ \\
      PreResNet-18 & CIFAR-100  & 2.00 & 75.8 & 75.8 & 7.30$\times$ \\
      MobileNet-v2 & CIFAR-10   & 3.17 & 93.8 & 94.2 & 4.33$\times$ \\
      MobileNet-v2 & CIFAR-100  & 2.89 & 72.2 & 72.1 & 4.89$\times$ \\
      DenseNet-121 & CIFAR-10   & 3.97 & 94.3 & 94.3 & 3.93$\times$ \\
      DenseNet-121 & CIFAR-10   & 1.86 & 93.6 & 94.3 & 4.21$\times$ \\
      \bottomrule
    \end{tabular}%
  }
    \caption{Performance of \FWName{} CNNs on CPU Vector Units. Speedups are computed against INT8 models. Bits per parameter (bpp) are also listed. \FWName{} consistently yields notable speedups over INT8 while preserving with full-precision accuracies.}

    % \vspace{-0.5cm}
    \label{tab:cnn_latency_cpu}
\end{table}

% \medskip

% \vspace{-0.5em}
\begin{figure}[ht]
    \includegraphics[width=0.95\linewidth]{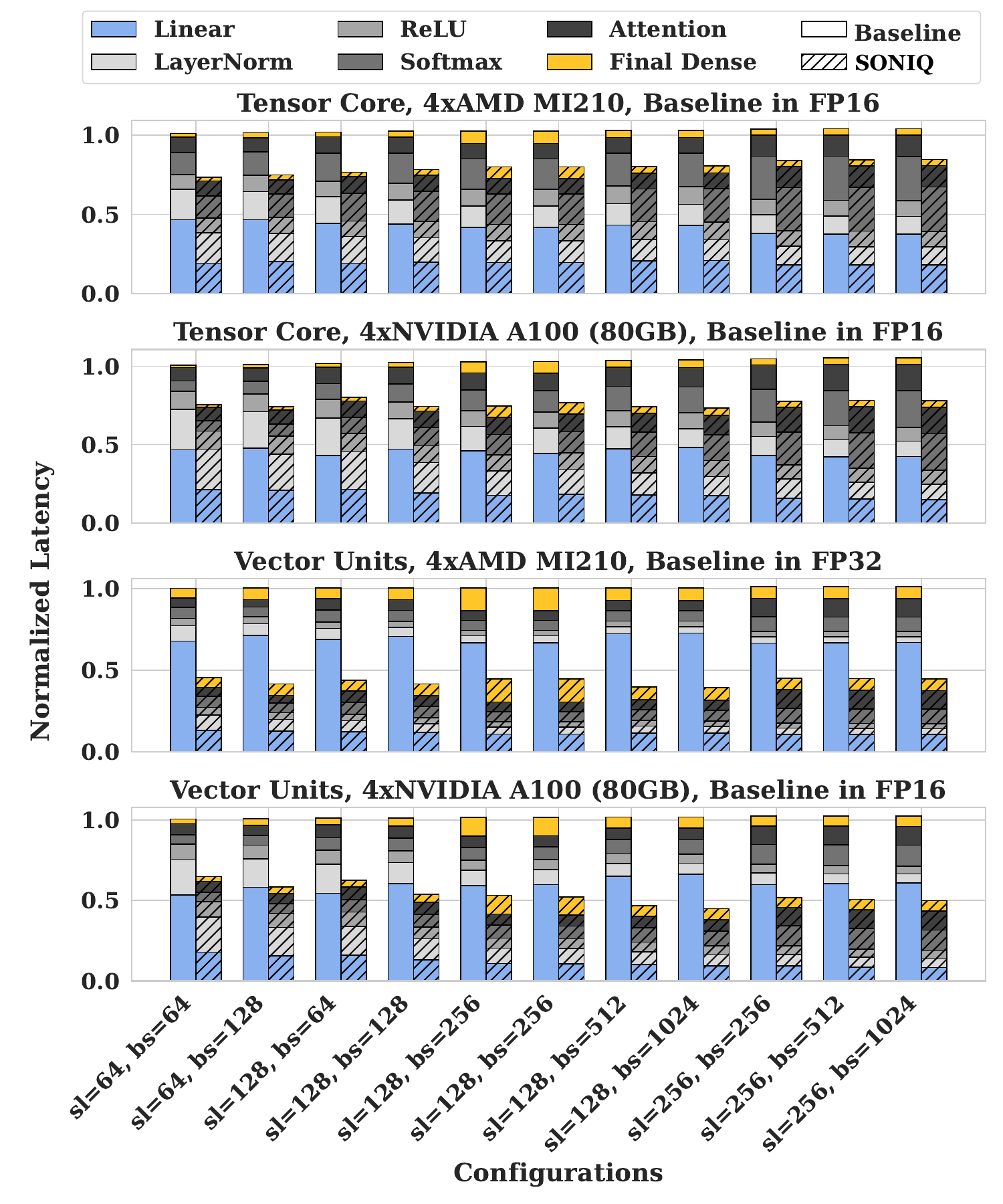}
    % \vspace{-0.3cm}
\caption{Inference latency breakdown of the transformer trained with \FWName{}. Here, $bs$ denotes batch size and $sl$ the (equal) source- and target-sequence length. We apply weight–activation quantization to “Linear’’ (all non-attention matmul layers except the final projection) and weight-only quantization to the final dense layer; greyed operators remain unmodified.}
    \label{fig:transformer_latency}
\end{figure}

\noindent\textbf{GPU results.} Figure~\ref{fig:transformer_latency} shows similarly substantial gains on GPUs. On vector units, \FWName{} is able to translate its compression ratio advantage into \textbf{7.3$\times$} and \textbf{6.2$\times$} speedups versus NVIDIA~A100’s FP16 and AMD~MI210’s FP32 implementations, respectively, when weight–activation quantization is applied. End-to-end, these benefits yield overall speedups of \textbf{1.5$\times$–2.5$\times$} and \textbf{2.3$\times$–2.8$\times$}. On tensor cores, whose architectures are already highly optimized for matrix multiplications, \FWName{} still delivers up to \textbf{3$\times$} speedups on weight–activation layers and an end-to-end improvement of up to \textbf{45\%}. These results are notable given that non-MatMul operators such as \texttt{Softmax} and \texttt{LayerNorm} contribute significantly to transformer latency. Finally, weight-only quantizing the last dense layer (highlighted in yellow) offers negligible additional latency gains, confirming that \FWName{} is already operating at the latency ceiling while outperforming the full-precision baseline in terms of task performance (i.e., BLEU score).

% highlighting the importance of joint weight–activation quantization—used by \FWName{} in most layers—for achieving substantial neural network acceleration.

% \yl{text here summarizing the results in the figure. point out trends, etc. If we can claim that the speed up roughly correspond to the compression ratio, that's a very strong result, as it shows that the networks are actually really hardware friendly.}

% \begin{table}[htbp]
% \centering
% \small
% \begin{tabular}{l l c c c c}
% \toprule
% \textbf{Architecture} & \textbf{Dataset} & \textbf{Bpp.} & \textbf{Acc.} & \textbf{FP Acc.} & \textbf{Speedup} \\
% \midrule
% PreResNet-18     & ImageNet & 6.25 & 70.8 & 69.6 & 2.87x \\
% PreResNet-18  & CIFAR-10 & 2.23 & 95.3 & 95.1 & 4.87x \\
% PreResNet-18  & CIFAR-10 & 1.98 & 95.0 & 95.1 & 7.15x \\
% PreResNet-18  & CIFAR-100 & 2.00 & 75.8 & 75.8 & 7.30x \\
% MobileNetv2 & CIFAR-10 & 3.17 & 93.8 & 94.2 & 4.33x \\
% MobileNetv2 & CIFAR-100 & 2.89 & 72.2 & 72.1 & 4.89x \\
% DenseNet-121  & CIFAR-10 & 3.97 & 94.3 & 94.3 & 3.93x \\
% DenseNet-121  & CIFAR-10 & 1.86 & 93.6 & 94.3 & 4.21x \\
% \bottomrule
% \end{tabular}

% \caption{Performance of \FWName{} CNNs on CPU Vector Units. Speedups computed against INT8. Bits per parameter (bpp) listed as well.}
% % \vspace{-0.5cm}
% \label{tab:cnn_latency_cpu}
% \end{table}

% \input{text/restrict_precisions}

% \input{text/related_work}
\section{Conclusion}\label{sec:conclusion}

\FWName{} demonstrates that \emph{system awareness is \textbf{the} missing ingredient for next-generation model compression}.  
By embedding hardware-calibrated perturbations directly into training, \FWName{-QAT} aligns algorithmic objectives with real deployment costs—latency, energy, and memory—while retaining fine-grained control over precision.  
Across CNNs and transformers, \FWName{} delivers up to 16× weight-activation compression, single-digit–millisecond inference on edge devices, and 7× CPU / 6× GPU throughput gains, all \emph{without sacrificing—and often surpassing—full-precision accuracy}.  
These results close the long-standing gap between mixed-precision research and commodity hardware support, proving that principled, noise-injection-based QAT can meet the exacting demands of mobile AR/VR, data-center scale, and safety-critical applications alike. System-aware quantization, such as \FWName{}, is a cornerstone of next-generation AI applications and technology.

Looking forward, we envision three avenues for broadening \FWName{}’s impact. \textbf{\textit{(1)}} extend \FWName{-QAT} to model hardware–induced noise—e.g., quantization artifacts in analog or optical accelerators—so training anticipates real-world signal distortions; \textbf{\textit{(2)}} integrating fairness and robustness constraints to ensure that efficiency never comes at the expense of vulnerable sub-populations; and \textbf{\textit{(3)}} co-optimizing quantization with neural architecture search to discover designs intrinsically amenable to ultra-low-bit deployment.

% \FWName{} matches or even exceeds full-precision accuracies while achieving compression ratios of up to 16x for CNNs and 7x for transformers. Furthermore, we illustrate the ease of fitting our design into existing hardware, using tensor cores and vector units on GPUs and vector units on CPUs as examples. From our experiments, we demonstrate speedups of up to 7.3x for CNNs on CPU vector units compared to INT8 implementations, and up to 6.3x and 2.8x for transformers on GPU vector units and tensor cores compared to FP16 baselines, respectively.

% In future work, we aim to develop systems that capture more diverse precision structures, improving the accuracy to bits-per-parameter trade-off, and explore post-training quantization approaches (i.e., quantization on pre-trained checkpoints) to reduce related training costs. Through our rigorous evaluation and exploration, the insights obtained and lessons learned will guide future research to push the frontiers of both network accuracy and system efficiency. We plan to open source our framework to facilitate further research endeavors, and extend our study to cover other hardware/system platforms and other co-design problems at the intersection of computer architecture and machine learning.


\bibliography{refs}

@article{liang2021pruning,
  title={Pruning and quantization for deep neural network acceleration: A survey},
  author={Liang, Tailin and Glossner, John and Wang, Lei and Shi, Shaobo and Zhang, Xiaotong},
  journal={Neurocomputing},
  volume={461},
  pages={370--403},
  year={2021},
  publisher={Elsevier}
}

@inproceedings{huang2017densely,
  title={Densely connected convolutional networks},
  author={Huang, Gao and Liu, Zhuang and Van Der Maaten, Laurens and Weinberger, Kilian Q},
  booktitle={Proceedings of the IEEE conference on computer vision and pattern recognition},
  pages={4700--4708},
  year={2017}
}

@inproceedings{sandler2018mobilenetv2,
  title={Mobilenetv2: Inverted residuals and linear bottlenecks},
  author={Sandler, Mark and Howard, Andrew and Zhu, Menglong and Zhmoginov, Andrey and Chen, Liang-Chieh},
  booktitle={Proceedings of the IEEE conference on computer vision and pattern recognition},
  pages={4510--4520},
  year={2018}
}

@inproceedings{he2016deep,
  title={Deep residual learning for image recognition},
  author={He, Kaiming and Zhang, Xiangyu and Ren, Shaoqing and Sun, Jian},
  booktitle={Proceedings of the IEEE conference on computer vision and pattern recognition},
  pages={770--778},
  year={2016}
}

@article{yang2024harnessing,
  title={Harnessing the power of llms in practice: A survey on chatgpt and beyond},
  author={Yang, Jingfeng and Jin, Hongye and Tang, Ruixiang and Han, Xiaotian and Feng, Qizhang and Jiang, Haoming and Zhong, Shaochen and Yin, Bing and Hu, Xia},
  journal={ACM Transactions on Knowledge Discovery from Data},
  volume={18},
  number={6},
  pages={1--32},
  year={2024},
  publisher={ACM New York, NY}
}

@article{street2024llms,
  title={LLMs achieve adult human performance on higher-order theory of mind tasks},
  author={Street, Winnie and Siy, John Oliver and Keeling, Geoff and Baranes, Adrien and Barnett, Benjamin and McKibben, Michael and Kanyere, Tatenda and Lentz, Alison and Dunbar, Robin IM and others},
  journal={arXiv preprint arXiv:2405.18870},
  year={2024}
}

@inproceedings{zhao2023more,
  title={More human than human: LLM-generated narratives outperform human-LLM interleaved narratives},
  author={Zhao, Zoie and Song, Sophie and Duah, Bridget and Macbeth, Jamie and Carter, Scott and Van, Monica P and Bravo, Nayeli Suseth and Klenk, Matthew and Sick, Kate and Filipowicz, Alexandre LS},
  booktitle={Proceedings of the 15th Conference on Creativity and Cognition},
  pages={368--370},
  year={2023}
}

@article{savarese2022not,
  title={Not all bits have equal value: Heterogeneous precisions via trainable noise},
  author={Savarese, Pedro and Yuan, Xin and Li, Yanjing and Maire, Michael},
  journal={Advances in Neural Information Processing Systems},
  volume={35},
  pages={35769--35782},
  year={2022}
}

@article{zhao2024atom,
  title={Atom: Low-bit quantization for efficient and accurate llm serving},
  author={Zhao, Yilong and Lin, Chien-Yu and Zhu, Kan and Ye, Zihao and Chen, Lequn and Zheng, Size and Ceze, Luis and Krishnamurthy, Arvind and Chen, Tianqi and Kasikci, Baris},
  journal={Proceedings of Machine Learning and Systems},
  volume={6},
  pages={196--209},
  year={2024}
}

@inproceedings{chauhan2023post,
  title={Post Training Mixed Precision Quantization of Neural Networks Using First-Order Information},
  author={Chauhan, Arun and Tiwari, Utsav and others},
  booktitle={Proceedings of the IEEE/CVF International Conference on Computer Vision},
  pages={1343--1352},
  year={2023}
}

@article{wang2022optimization,
  title={Optimization-based post-training quantization with bit-split and stitching},
  author={Wang, Peisong and Chen, Weihan and He, Xiangyu and Chen, Qiang and Liu, Qingshan and Cheng, Jian},
  journal={IEEE Transactions on Pattern Analysis and Machine Intelligence},
  volume={45},
  number={2},
  pages={2119--2135},
  year={2022},
  publisher={IEEE}
}

@inproceedings{gong2019differentiable,
  title={Differentiable soft quantization: Bridging full-precision and low-bit neural networks},
  author={Gong, Ruihao and Liu, Xianglong and Jiang, Shenghu and Li, Tianxiang and Hu, Peng and Lin, Jiazhen and Yu, Fengwei and Yan, Junjie},
  booktitle={Proceedings of the IEEE/CVF international conference on computer vision},
  pages={4852--4861},
  year={2019}
}

@article{choi2018pact,
  title={Pact: Parameterized clipping activation for quantized neural networks},
  author={Choi, Jungwook and Wang, Zhuo and Venkataramani, Swagath and Chuang, Pierce I-Jen and Srinivasan, Vijayalakshmi and Gopalakrishnan, Kailash},
  journal={arXiv preprint arXiv:1805.06085},
  year={2018}
}

@article{yang2021bsq,
  title={BSQ: Exploring bit-level sparsity for mixed-precision neural network quantization},
  author={Yang, Huanrui and Duan, Lin and Chen, Yiran and Li, Hai},
  journal={arXiv preprint arXiv:2102.10462},
  year={2021}
}

@article{yin2018binaryrelax,
  title={Binaryrelax: A relaxation approach for training deep neural networks with quantized weights},
  author={Yin, Penghang and Zhang, Shuai and Lyu, Jiancheng and Osher, Stanley and Qi, Yingyong and Xin, Jack},
  journal={SIAM Journal on Imaging Sciences},
  volume={11},
  number={4},
  pages={2205--2223},
  year={2018},
  publisher={SIAM}
}

@inproceedings{liu2021layer,
  title={Layer importance estimation with imprinting for neural network quantization},
  author={Liu, Hongyang and Elkerdawy, Sara and Ray, Nilanjan and Elhoushi, Mostafa},
  booktitle={Proceedings of the IEEE/CVF Conference on Computer Vision and Pattern Recognition},
  pages={2408--2417},
  year={2021}
}

@inproceedings{lee2022toward,
  title={Toward efficient low-precision training: Data format optimization and hysteresis quantization},
  author={Lee, Sunwoo and Park, Jeongwoo and Jeon, Dongsuk},
  booktitle={International Conference on Learning Representations},
  year={2022}
}

@inproceedings{guo2023olive,
  title={Olive: Accelerating large language models via hardware-friendly outlier-victim pair quantization},
  author={Guo, Cong and Tang, Jiaming and Hu, Weiming and Leng, Jingwen and Zhang, Chen and Yang, Fan and Liu, Yunxin and Guo, Minyi and Zhu, Yuhao},
  booktitle={Proceedings of the 50th Annual International Symposium on Computer Architecture},
  pages={1--15},
  year={2023}
}

@inproceedings{patsidis2020risc,
  title={RISC-V 2: a scalable RISC-V vector processor},
  author={Patsidis, Kariofyllis and Nicopoulos, Chrysostomos and Sirakoulis, Georgios Ch and Dimitrakopoulos, Giorgos},
  booktitle={2020 IEEE International Symposium on Circuits and Systems (ISCAS)},
  pages={1--5},
  year={2020},
  organization={IEEE}
}

@article{zhang2023hardware,
  title={A Hardware Evaluation Framework for Large Language Model Inference},
  author={Zhang, Hengrui and Ning, August and Prabhakar, Rohan and Wentzlaff, David},
  journal={arXiv preprint arXiv:2312.03134},
  year={2023}
}

@inproceedings{zhong2022dynamic,
  title={Dynamic dual trainable bounds for ultra-low precision super-resolution networks},
  author={Zhong, Yunshan and Lin, Mingbao and Li, Xunchao and Li, Ke and Shen, Yunhang and Chao, Fei and Wu, Yongjian and Ji, Rongrong},
  booktitle={European Conference on Computer Vision},
  pages={1--18},
  year={2022},
  organization={Springer}
}

@article{liu2023ultra,
  title={Ultra-low Precision Multiplication-free Training for Deep Neural Networks},
  author={Liu, Chang and Zhang, Rui and Zhang, Xishan and Hao, Yifan and Du, Zidong and Hu, Xing and Li, Ling and Guo, Qi},
  journal={arXiv preprint arXiv:2302.14458},
  year={2023}
}

@inproceedings{tang2022mixed,
  title={Mixed-Precision Neural Network Quantization via Learned Layer-Wise Importance},
  author={Tang, Chen and Ouyang, Kai and Wang, Zhi and Zhu, Yifei and Ji, Wen and Wang, Yaowei and Zhu, Wenwu},
  booktitle={European Conference on Computer Vision},
  pages={259--275},
  year={2022},
  organization={Springer}
}

@inproceedings{wang2019haq,
  title={Haq: Hardware-aware automated quantization with mixed precision},
  author={Wang, Kuan and Liu, Zhijian and Lin, Yujun and Lin, Ji and Han, Song},
  booktitle={Proceedings of the IEEE/CVF conference on computer vision and pattern recognition},
  pages={8612--8620},
  year={2019}
}

@article{nagel2021white,
  title={A white paper on neural network quantization},
  author={Nagel, Markus and Fournarakis, Marios and Amjad, Rana Ali and Bondarenko, Yelysei and Van Baalen, Mart and Blankevoort, Tijmen},
  journal={arXiv preprint arXiv:2106.08295},
  year={2021}
}

@article{choi2019accurate,
  title={Accurate and efficient 2-bit quantized neural networks},
  author={Choi, Jungwook and Venkataramani, Swagath and Srinivasan, Vijayalakshmi Viji and Gopalakrishnan, Kailash and Wang, Zhuo and Chuang, Pierce},
  journal={Proceedings of Machine Learning and Systems},
  volume={1},
  pages={348--359},
  year={2019}
}

@inproceedings{chen2018tvm,
  title={$\{$TVM$\}$: An automated $\{$End-to-End$\}$ optimizing compiler for deep learning},
  author={Chen, Tianqi and Moreau, Thierry and Jiang, Ziheng and Zheng, Lianmin and Yan, Eddie and Shen, Haichen and Cowan, Meghan and Wang, Leyuan and Hu, Yuwei and Ceze, Luis and others},
  booktitle={13th USENIX Symposium on Operating Systems Design and Implementation (OSDI 18)},
  pages={578--594},
  year={2018}
}

@article{esser2019learned,
  title={Learned step size quantization},
  author={Esser, Steven K and McKinstry, Jeffrey L and Bablani, Deepika and Appuswamy, Rathinakumar and Modha, Dharmendra S},
  journal={arXiv preprint arXiv:1902.08153},
  year={2019}
}

@article{fan2020training,
  title={Training with quantization noise for extreme model compression},
  author={Fan, Angela and Stock, Pierre and Graham, Benjamin and Grave, Edouard and Gribonval, R{\'e}mi and Jegou, Herve and Joulin, Armand},
  journal={arXiv preprint arXiv:2004.07320},
  year={2020}
}

@article{ibrahim2012greda,
  title={GREDA: A fast and more accurate gate reliability EDA tool},
  author={Ibrahim, Walid and Beiu, Valeriu and Beg, Azam},
  journal={IEEE Transactions on Computer-Aided Design of Integrated Circuits and Systems},
  volume={31},
  number={4},
  pages={509--521},
  year={2012},
  publisher={IEEE}
}

@misc{nvidia_ampere,
  author = {NVIDIA},
  title = {NVIDIA Ampere Architecture In-Depth},
  year = {2020},
  month = {May},
  howpublished = {\url{https://developer.nvidia.com/blog/nvidia-ampere-architecture-in-depth/}},
  note = {Accessed: 2024-05-05}
}

@article{zhao2023atom,
  title={Atom: Low-bit quantization for efficient and accurate llm serving},
  author={Zhao, Yilong and Lin, Chien-Yu and Zhu, Kan and Ye, Zihao and Chen, Lequn and Zheng, Size and Ceze, Luis and Krishnamurthy, Arvind and Chen, Tianqi and Kasikci, Baris},
  journal={arXiv preprint arXiv:2310.19102},
  year={2023}
}

@inproceedings{liu2019optimizing,
  title={Optimizing $\{$CNN$\}$ model inference on $\{$CPUs$\}$},
  author={Liu, Yizhi and Wang, Yao and Yu, Ruofei and Li, Mu and Sharma, Vin and Wang, Yida},
  booktitle={2019 USENIX Annual Technical Conference (USENIX ATC 19)},
  pages={1025--1040},
  year={2019}
}

@inproceedings{zhang2018shufflenet,
  title={Shufflenet: An extremely efficient convolutional neural network for mobile devices},
  author={Zhang, Xiangyu and Zhou, Xinyu and Lin, Mengxiao and Sun, Jian},
  booktitle={Proceedings of the IEEE conference on computer vision and pattern recognition},
  pages={6848--6856},
  year={2018}
}

@article{krizhevsky2009learning,
  title={Learning Multiple Layers of Features from Tiny Images},
  author={Krizhevsky, Alex},
  year={2009},
  publisher={Citeseer}
}

@inproceedings{deng2009imagenet,
  title={Imagenet: A large-scale hierarchical image database},
  author={Deng, Jia and Dong, Wei and Socher, Richard and Li, Li-Jia and Li, Kai and Fei-Fei, Li},
  booktitle={2009 IEEE conference on computer vision and pattern recognition},
  pages={248--255},
  year={2009},
  organization={Ieee}
}

@article{binkert2011gem5,
  title={The gem5 simulator},
  author={Binkert, Nathan and Beckmann, Bradford and Black, Gabriel and Reinhardt, Steven K and Saidi, Ali and Basu, Arkaprava and Hestness, Joel and Hower, Derek R and Krishna, Tushar and Sardashti, Somayeh and others},
  journal={ACM SIGARCH computer architecture news},
  volume={39},
  number={2},
  pages={1--7},
  year={2011},
  publisher={ACM New York, NY, USA}
}

@misc{lec03_gpu_architectures,
  title        = {Lecture 3: GPU Architectures},
  author       = {{European Commission}},
  howpublished = {\url{https://ec.europa.eu/programmes/erasmus-plus/project-result-content/52dfac24-28e9-4379-8f28-f8ed05e225e0/lec03\_gpu\_architectures.pdf}},
  note         = {Accessed on: 2024-05-02}
}

@inproceedings{zhou2023simd,
  title={YFlows: Systematic Dataflow Exploration and Code Generation for Efficient Neural Network Inference using SIMD Architectures on CPUs},
  author={Zhou, Cyrus and Hassman, Zack and Shah, Dhirpal and Richard, Vaughn and Li, Yanjing},
  booktitle={Proceedings of the 33rd ACM SIGPLAN International Conference on Compiler Construction},
  pages={212--226},
  year={2024}
}

@article{chen2019eyeriss,
  title={Eyeriss v2: A flexible accelerator for emerging deep neural networks on mobile devices},
  author={Chen, Yu-Hsin and Yang, Tien-Ju and Emer, Joel and Sze, Vivienne},
  journal={IEEE Journal on Emerging and Selected Topics in Circuits and Systems},
  volume={9},
  number={2},
  pages={292--308},
  year={2019},
  publisher={IEEE}
}

@article{lin2024awq,
  title={AWQ: Activation-aware Weight Quantization for On-Device LLM Compression and Acceleration},
  author={Lin, Ji and Tang, Jiaming and Tang, Haotian and Yang, Shang and Chen, Wei-Ming and Wang, Wei-Chen and Xiao, Guangxuan and Dang, Xingyu and Gan, Chuang and Han, Song},
  journal={Proceedings of Machine Learning and Systems},
  volume={6},
  pages={87--100},
  year={2024}
}

@article{zhao2023recd,
  title={RecD: Deduplication for end-to-end deep learning recommendation model training infrastructure},
  author={Zhao, Mark and Choudhary, Dhruv and Tyagi, Devashish and Somani, Ajay and Kaplan, Max and Lin, Sung-Han and Pumma, Sarunya and Park, Jongsoo and Basant, Aarti and Agarwal, Niket and others},
  journal={Proceedings of Machine Learning and Systems},
  volume={5},
  pages={754--767},
  year={2023}
}

@inproceedings{hazelwood2018applied,
  title={Applied machine learning at facebook: A datacenter infrastructure perspective},
  author={Hazelwood, Kim and Bird, Sarah and Brooks, David and Chintala, Soumith and Diril, Utku and Dzhulgakov, Dmytro and Fawzy, Mohamed and Jia, Bill and Jia, Yangqing and Kalro, Aditya and others},
  booktitle={2018 IEEE international symposium on high performance computer architecture (HPCA)},
  pages={620--629},
  year={2018},
  organization={IEEE}
}

@inproceedings{nguyen2024s,
  title={S. Towards sustainable large language model serving},
  author={Nguyen, Sophia and Zhou, Beihao and Liu, YD},
  booktitle={ACM HotCarbon Workshop on Sustainable Computer Systems},
  year={2024}
}

@article{mcdonald2022great,
  title={Great power, great responsibility: Recommendations for reducing energy for training language models},
  author={McDonald, Joseph and Li, Baolin and Frey, Nathan and Tiwari, Devesh and Gadepally, Vijay and Samsi, Siddharth},
  journal={arXiv preprint arXiv:2205.09646},
  year={2022}
}

@article{zhu2024nanoflow,
  title={NanoFlow: Towards Optimal Large Language Model Serving Throughput},
  author={Zhu, Kan and Zhao, Yilong and Zhao, Liangyu and Zuo, Gefei and Gu, Yile and Xie, Dedong and Gao, Yufei and Xu, Qinyu and Tang, Tian and Ye, Zihao and others},
  journal={arXiv preprint arXiv:2408.12757},
  year={2024}
}

@article{oliaro2024suffixdecoding,
  title={SuffixDecoding: A Model-Free Approach to Speeding Up Large Language Model Inference},
  author={Oliaro, Gabriele and Jia, Zhihao and Campos, Daniel and Qiao, Aurick},
  journal={arXiv preprint arXiv:2411.04975},
  year={2024}
}

@article{brown2024large,
  title={Large language monkeys: Scaling inference compute with repeated sampling},
  author={Brown, Bradley and Juravsky, Jordan and Ehrlich, Ryan and Clark, Ronald and Le, Quoc V and R{\'e}, Christopher and Mirhoseini, Azalia},
  journal={arXiv preprint arXiv:2407.21787},
  year={2024}
}

@inproceedings{wang2022learnable,
  title={Learnable lookup table for neural network quantization},
  author={Wang, Longguang and Dong, Xiaoyu and Wang, Yingqian and Liu, Li and An, Wei and Guo, Yulan},
  booktitle={Proceedings of the IEEE/CVF conference on computer vision and pattern recognition},
  pages={12423--12433},
  year={2022}
}

@misc{intel_avx512,
  title = {Intel® Advanced Vector Extensions 512 (Intel® AVX-512) Overview},
  author = {Intel Corporation},
  year = {2024},
  url = {https://www.intel.com/content/www/us/en/architecture-and-technology/avx-512-overview.html},
  note = {Accessed: 2024-11-18}
}

@article{vaswani2017attention,
  title={Attention is all you need},
  author={Vaswani, A},
  journal={Advances in Neural Information Processing Systems},
  year={2017}
}

@article{croitoru2023diffusion,
  title={Diffusion models in vision: A survey},
  author={Croitoru, Florinel-Alin and Hondru, Vlad and Ionescu, Radu Tudor and Shah, Mubarak},
  journal={IEEE Transactions on Pattern Analysis and Machine Intelligence},
  volume={45},
  number={9},
  pages={10850--10869},
  year={2023},
  publisher={IEEE}
}

@article{zhou2016dorefa,
  title={Dorefa-net: Training low bitwidth convolutional neural networks with low bitwidth gradients},
  author={Zhou, Shuchang and Wu, Yuxin and Ni, Zekun and Zhou, Xinyu and Wen, He and Zou, Yuheng},
  journal={arXiv preprint arXiv:1606.06160},
  year={2016}
}

@inproceedings{cettolo2014report,
  title        = {Report on the 11th IWSLT Evaluation Campaign},
  author       = {Cettolo, Mauro and Niehues, Jan and Bentivogli, Luisa and Bertoldi, Nicola and Federico, Marcello},
  booktitle    = {Proceedings of the 11th International Workshop on Spoken Language Translation (IWSLT)},
  year         = {2014},
  pages        = {2--17},
  organization = {IWSLT},
}

@article{hoang2020direct,
  title={Direct quantization for training highly accurate low bit-width deep neural networks},
  author={Hoang, Tuan and Do, Thanh-Toan and Nguyen, Tam V and Cheung, Ngai-Man},
  journal={arXiv preprint arXiv:2012.13762},
  year={2020}
}

@inproceedings{romero2021infaas,
  title={$\{$INFaaS$\}$: Automated model-less inference serving},
  author={Romero, Francisco and Li, Qian and Yadwadkar, Neeraja J and Kozyrakis, Christos},
  booktitle={2021 USENIX Annual Technical Conference (USENIX ATC 21)},
  pages={397--411},
  year={2021}
}

@article{patterson2024energy,
  title={Energy and emissions of machine learning on smartphones vs. the cloud},
  author={Patterson, David and Gilbert, Jeffrey M and Gruteser, Marco and Robles, Efren and Sekar, Krishna and Wei, Yong and Zhu, Tenghui},
  journal={Communications of the ACM},
  volume={67},
  number={2},
  pages={86--97},
  year={2024},
  publisher={ACM New York, NY, USA}
}

@article{nigade2024inference,
  title={Inference serving with end-to-end latency SLOs over dynamic edge networks},
  author={Nigade, Vinod and Bauszat, Pablo and Bal, Henri and Wang, Lin},
  journal={Real-Time Systems},
  volume={60},
  number={2},
  pages={239--290},
  year={2024},
  publisher={Springer}
}

@article{hooker2020characterising,
  title={Characterising bias in compressed models},
  author={Hooker, Sara and Moorosi, Nyalleng and Clark, Gregory and Bengio, Samy and Denton, Emily},
  journal={arXiv preprint arXiv:2010.03058},
  year={2020}
}

@inproceedings{reddi2022mlperf,
  author    = {Vijay Janapa Reddi and David Kanter and Peter Mattson and Jared Duke and
               Thai Nguyen and \textit{et~al.}},
  title     = {{MLPerf} Mobile Inference Benchmark: An Industry-Standard Open-Source Machine Learning Benchmark for On-Device AI},
  booktitle = {Proceedings of the 5th Conference on Machine Learning and Systems (MLSys)},
  pages     = {352--369},
  year      = {2022}
}

@phdthesis{lin2017latency,
  author  = {Lincoln, Jesse},
  title   = {Low Latency Displays for Augmented Reality},
  school  = {University of Central Florida},
  year    = {2017}
}

@article{lin2022ondevice,
  author  = {Lin, Ji and Zhu, Ligeng and Chen, Wei-Ming and Wang, Wei-Chen and Gan, Chuang and Han, Song},
  title   = {On-Device Training Under 256 KB Memory},
  journal = {arXiv preprint arXiv:2206.15472},
  year    = {2022}
}

@article{barbosa2025af,
  author  = {Barbosa, Isabella O. F. and Oliveira, Beatriz C. and Santos, Charles K. M. and Miranda, Maria C. R. and Barbosa, Gabriel A. and Menezes~Jr., Antônio S.},
  title   = {Smartphone-Based Applications for Atrial Fibrillation Detection: A Systematic Review and Meta-Analysis of Diagnostic Test Accuracy},
  journal = {Telemedicine and e-Health},
  year    = {2025},
  doi     = {10.1089/tmj.2024.0579}
}

@inproceedings{navardi2022drone,
  author    = {Navardi, Mozhgan and Shiri, Aidin and Humes, Edward and Waytowich, Nicholas R. and Mohsenin, Tinoosh},
  title     = {An Optimization Framework for Efficient Vision-Based Autonomous Drone Navigation},
  booktitle = {IEEE International Conference on Artificial Intelligence Circuits and Systems (AICAS)},
  pages     = {304--307},
  year      = {2022},
  doi       = {10.1109/AICAS54582.2022.9817444}
}

@article{zhou2025lowra,
  title={LowRA: Accurate and Efficient LoRA Fine-Tuning of LLMs under 2 Bits},
  author={Zhou, Zikai and Zhang, Qizheng and Kumbong, Hermann and Olukotun, Kunle},
  journal={arXiv preprint arXiv:2502.08141},
  year={2025}
}

@inproceedings{dong2023emq,
  title={Emq: Evolving training-free proxies for automated mixed precision quantization},
  author={Dong, Peijie and Li, Lujun and Wei, Zimian and Niu, Xin and Tian, Zhiliang and Pan, Hengyue},
  booktitle={Proceedings of the IEEE/CVF international conference on computer vision},
  pages={17076--17086},
  year={2023}
}

@inproceedings{fan2020quantnoise,
  author    = {Angela Fan and David Grangier and Michael Auli},
  title     = {Training with Quantization Noise for Extreme Model Compression},
  booktitle = {International Conference on Learning Representations (ICLR)},
  year      = {2021}
}

@article{bishop1995training,
  author  = {Christopher M. Bishop},
  title   = {Training with Noise is Equivalent to Tikhonov Regularization},
  journal = {Neural Computation},
  volume  = {7},
  number  = {1},
  pages   = {108--116},
  year    = {1995}
}

@article{srivastava2014dropout,
  author  = {Nitish Srivastava and Geoffrey Hinton and Alex Krizhevsky and Ilya Sutskever and Ruslan Salakhutdinov},
  title   = {Dropout: A Simple Way to Prevent Neural Networks from Overfitting},
  journal = {Journal of Machine Learning Research},
  volume  = {15},
  pages   = {1929--1958},
  year    = {2014}
}

@inproceedings{fortunato2017noisy,
  author    = {Meire Fortunato and Mohammad Gheshlaghi Azar and Bilal Piot and et~al.},
  title     = {Noisy Networks for Exploration},
  booktitle = {International Conference on Learning Representations (ICLR)},
  year      = {2018}
}

@inproceedings{madry2018towards,
  author    = {Aleksander Madry and Aleksandar Makelov and Ludwig Schmidt and Dimitris Tsipras and Adrian Vladu},
  title     = {Towards Deep Learning Models Resistant to Adversarial Attacks},
  booktitle = {International Conference on Learning Representations (ICLR)},
  year      = {2018}
}

@inproceedings{vincent2008extracting,
  author    = {Pascal Vincent and Hugo Larochelle and Yoshua Bengio and Pierre-Antoine Manzagol},
  title     = {Extracting and Composing Robust Features with Denoising Autoencoders},
  booktitle = {International Conference on Machine Learning (ICML)},
  pages     = {1096--1103},
  year      = {2008}
}

@manual{intel_vnni,
  title        = {{Intel}{\textregistered} 64 and IA-32 Architectures Software Developer’s Manual, Vol.\ 2A: Instruction Set Reference, A–M},
  subtitle     = {Section 5.17 “AVX512 Vector Neural Network Instructions (VNNI)”},
  author       = {{Intel Corporation}},
  year         = {2023},
  note         = {Doc.\ no.\ 325383-076},
  url          = {https://www.intel.com/content/www/us/en/developer/articles/technical/intel-sdm.html}
}

@manual{amd_zen4_vnni,
  title        = {{AMD64} Architecture Programmer’s Manual, Volume 2: System Programming},
  subtitle     = {§7.10 “AVX-512 Vector Neural Network Instructions”},
  author       = {{Advanced Micro Devices, Inc.}},
  year         = {2023},
  url          = {https://developer.amd.com/resources/developer-guides-manuals/}
}

@manual{arm_usdot,
  title        = {Arm{\textregistered} Architecture Reference Manual, Armv8{-}A},
  subtitle     = {DDI 0487H.a, Section G7.256 “UDOT/SDOT — Dot-product of 8-bit integers”},
  author       = {{Arm Ltd.}},
  year         = {2022},
  url          = {https://developer.arm.com/documentation/ddi0487/latest}
}

@misc{nvidia_tc_int8,
  title        = {NVIDIA Turing GPU Architecture},
  howpublished = {White Paper},
  author       = {{NVIDIA Corporation}},
  year         = {2018},
  note         = {See §6.2 “Integer Tensor Core Operations (INT8, INT4)”},
  url          = {https://www.nvidia.com/en-us/geforce/turing/}
}

@misc{nvidia_tc_int4,
  title        = {NVIDIA Ampere GPU Architecture},
  howpublished = {White Paper},
  author       = {{NVIDIA Corporation}},
  year         = {2020},
  note         = {§5.3 “Tensor Cores and Low-Precision Modes (INT8/INT4)”},
  url          = {https://resources.nvidia.com/en-us-turing-architecture}
}

@misc{google_edgetpu,
  title        = {Coral {Edge TPU} Datasheet},
  howpublished = {Tech.\ datasheet},
  author       = {{Google LLC}},
  year         = {2021},
  note         = {“The Edge TPU executes 8-bit unsigned activations $\times$ 8-bit signed weights with INT32 accumulation.”},
  url          = {https://coral.ai/docs/edgetpu/datasheet/}
}

@article{jang2016categorical,
  title={Categorical reparameterization with gumbel-softmax},
  author={Jang, Eric and Gu, Shixiang and Poole, Ben},
  journal={arXiv preprint arXiv:1611.01144},
  year={2016}
}

@inproceedings{xiao2023smoothquant,
  author    = {Guangxuan Xiao and Ji Lin and Mickael Seznec and Hao Wu and Julien Demouth and Song Han},
  title     = {SmoothQuant: Accurate and Efficient Post-Training Quantization for Large Language Models},
  booktitle = {International Conference on Machine Learning (ICML)},
  year      = {2023},
  url       = {https://arxiv.org/abs/2211.10438}
}

@article{courbariaux2015binaryconnect,
  title         = {BinaryConnect: Training Deep Neural Networks with Binary Weights during Propagations},
  author        = {Courbariaux, Matthieu and Bengio, Yoshua and David, Jean-Pierre},
  journal       = {Advances in Neural Information Processing Systems (NeurIPS) Workshop},
  year          = {2015},
  note          = {arXiv:1511.00363}
}

@inproceedings{courbariaux2016binarynet,
  title         = {BinaryNet: Training Deep Neural Networks with Weights and Activations Constrained to +1 or -1},
  author        = {Courbariaux, Matthieu and Hubara, Itay and Soudry, Daniel and El-Yaniv, Ran and Bengio, Yoshua},
  booktitle     = {Advances in Neural Information Processing Systems (NeurIPS)},
  year          = {2016}
}

@inproceedings{rastegari2016xnor,
  title         = {XNOR-Net: ImageNet Classification Using Binary Convolutional Neural Networks},
  author        = {Rastegari, Mohammad and Ordonez, Vicente and Redmon, Joseph and Farhadi, Ali},
  booktitle     = {European Conference on Computer Vision (ECCV)},
  year          = {2016},
  pages         = {525--542}
}

@inproceedings{umuroglu2017finn,
  title         = {FINN: A Framework for Fast, Scalable Binarized Neural Network Inference},
  author        = {Umuroglu, Yaman and Fraser, Nicholas J. and Gambardella, Giulio et al.},
  booktitle     = {ACM/SIGDA International Symposium on Field-Programmable Gate Arrays (FPGA)},
  year          = {2017},
  pages         = {65--74}
}

@inproceedings{sharma2018bitfusion,
  title         = {BitFusion: Bit-Level Dynamically Composable Architecture for Accelerating Deep Neural Networks},
  author        = {Sharma, Hyesoon and Park, Woojin and Bae, Jungwook et al.},
  booktitle     = {International Symposium on Computer Architecture (ISCA)},
  year          = {2018},
  pages         = {764--775}
}

@misc{nvidia_turing_2018,
  author       = {{NVIDIA Corporation}},
  title        = {{NVIDIA Turing GPU Architecture Whitepaper}},
  year         = {2018},
  note         = {Mixed INT8/INT4 Tensor Cores for DL inference},
  url          = {https://images.nvidia.com/.../NVIDIA-Turing-Architecture-Whitepaper.pdf}
}

@misc{nvidia_ampere_2020,
  author       = {{NVIDIA Corporation}},
  title        = {{NVIDIA A100 Tensor Core GPU Architecture Whitepaper}},
  year         = {2020},
  note         = {Adds INT4 and INT1 support, extends mixed‑precision Tensor Cores},
  url          = {https://images.nvidia.com/.../nvidia-ampere-architecture-whitepaper.pdf}
}

@misc{nvidia_hopper_2022,
  author       = {{NVIDIA Corporation}},
  title        = {{NVIDIA H100 Tensor Core GPU Architecture Overview}},
  year         = {2022},
  howpublished = {\url{https://resources.nvidia.com/.../gtc22-whitepaper-hopper}},
  note         = {Introduces FP8/FP16–INT4 mixed‑precision “Transformer Engine”}
}

@inproceedings{jouppi_tpu_isca17,
  author       = {Norman P. Jouppi and Cliff Young and Nishant Patil and David Patterson and others},
  title        = {{In‑Datacenter Performance Analysis of a Tensor Processing Unit}},
  booktitle    = {Proc.\ ISCA ’17},
  year         = {2017},
  pages        = {1--12},
  doi          = {10.1145/3079856.3080246}
}

@article{jouppi_supercomputer_cacm20,
  author  = {Norman P. Jouppi and Doe Hyun Yoon and George Kurian and others},
  title   = {{A Domain‑Specific Supercomputer for Training Deep Neural Networks}},
  journal = {Communications of the ACM},
  volume  = {63},
  number  = {7},
  pages   = {67--78},
  year    = {2020},
  doi     = {10.1145/3360307}
}

@misc{jouppi_tpu_v4_2023,
  author       = {Norman P. Jouppi and George Kurian and Sheng Li and others},
  title        = {{TPU v4: An Optically Reconfigurable Supercomputer for Machine Learning}},
  year         = {2023},
  howpublished = {\url{https://arxiv.org/abs/2304.01433}}
}

@misc{amd_cdna2_2021,
  author       = {{Advanced Micro Devices, Inc.}},
  title        = {{AMD CDNA\textsuperscript{TM} 2 Architecture White Paper}},
  year         = {2021},
  note         = {Matrix Cores support INT8 and INT4},
  url          = {https://www.amd.com/.../amd-cdna2-white-paper.pdf}
}

@misc{apple_coreml_wwdc24,
  author       = {{Apple Inc.}},
  title        = {{Bring your machine‑learning models to Apple Silicon (WWDC 2024 Session 10159)}},
  year         = {2024},
  howpublished = {\url{https://developer.apple.com/videos/play/wwdc2024/10159/}},
  note         = {Announces 4‑bit post‑training quantization on Neural Engine}
}

@misc{intel_gaudi2_2023,
  author       = {{Intel Corporation}},
  title        = {{Accelerating Large Language Models with Habana Gaudi2}},
  year         = {2023},
  howpublished = {\url{https://github.com/huggingface/blog/blob/main/habana-gaudi-2-bloom.md}},
  note         = {Shows 4‑bit and 8‑bit inference on Gaudi 2}
}

@inproceedings{bitfusion_isca18,
  author       = {Hadi M. Esmaeilzadeh and others},
  title        = {{Bit Fusion: Bit‑Level Dynamically Composable Architecture for Accelerating Deep Neural Networks}},
  booktitle    = {Proc.\ ISCA ’18},
  year         = {2018},
  pages        = {461--472},
  doi          = {10.1109/ISCA.2018.00043}
}

@article{bengio2013estimating,
  title={Estimating or propagating gradients through stochastic neurons for conditional computation},
  author={Bengio, Yoshua and L{\'e}onard, Nicholas and Courville, Aaron},
  journal={arXiv preprint arXiv:1308.3432},
  year={2013}
}

@article{meuser2024revisiting,
  title={Revisiting edge ai: Opportunities and challenges},
  author={Meuser, Tobias and Lov{\'e}n, Lauri and Bhuyan, Monowar and Patil, Shishir G and Dustdar, Schahram and Aral, Atakan and Bayhan, Suzan and Becker, Christian and De Lara, Eyal and Ding, Aaron Yi and others},
  journal={IEEE Internet Computing},
  volume={28},
  number={4},
  pages={49--59},
  year={2024},
  publisher={IEEE}
}

@article{li2024investigating,
  title={Investigating the impact of quantization on adversarial robustness},
  author={Li, Qun and Meng, Yuan and Tang, Chen and Jiang, Jiacheng and Wang, Zhi},
  journal={arXiv preprint arXiv:2404.05639},
  year={2024}
}

@inproceedings{xu2023q,
  title={Q-detr: An efficient low-bit quantized detection transformer},
  author={Xu, Sheng and Li, Yanjing and Lin, Mingbao and Gao, Peng and Guo, Guodong and L{\"u}, Jinhu and Zhang, Baochang},
  booktitle={Proceedings of the IEEE/CVF Conference on Computer Vision and Pattern Recognition},
  pages={3842--3851},
  year={2023}
}
\bibliographystyle{mlsys2025}

\appendix    

\section{Novelty and Positioning} \label{app:novelty}
\label{sec:novelty}

\begin{table*}
  \vspace{1em}
  \centering
  \small
  \setlength{\tabcolsep}{6pt}
  \begin{tabular}{@{}lccccccc@{}}
    \toprule
    \textbf{Criterion} & \textbf{\FWName{}} & \textbf{HAQ} & \textbf{TQN} & \textbf{SMOL} & \textbf{Bit\mbox{-}Split} & \textbf{LSQ} & \textbf{BSQ} \\
    \midrule
    W/A$<\!8$b $>$ FP (Transformer)       & \cmark & \xmark & \xmark & \xmark & \xmark & \xmark & \xmark \\
    W/A$<\!8$b $\approx$ FP (Transformer) & \cmark & \xmark & \cmark & \cmark & \xmark & \xmark & \xmark \\
    W/A$<\!8$b $>$ FP (CNN)               & \cmark & \cmark & \xmark & \cmark & \xmark & \cmark & \xmark \\
    W/A$<\!8$b $\approx$ FP (CNN)         & \cmark & \cmark & \xmark & \cmark & \xmark & \cmark & \xmark \\
    Zero custom runtime                   & \cmark & \xmark & \xmark & \xmark & \xmark & \cmark & \xmark \\
    Commodity xPU-ready                   & \cmark & \xmark & \xmark & \xmark & \xmark & \cmark & \xmark \\
    Realistic implementation exists       & \cmark & \cmark & \cmark & \xmark & \cmark & \cmark & \cmark \\
    \bottomrule
  \end{tabular}
  \caption{Positioning \FWName{} against representative baselines. ``W/A'' denotes weight/activation bit-width.}
  \label{tab:\FWName{}-comparison-mlsys}
  \vspace{2em}
\end{table*}

\textbf{Claim (to our knowledge).}
\emph{\FWName{}} is the first quantization framework that simultaneously:
(i) matches or exceeds full-precision (FP) accuracy on both Transformers and CNNs at sub-8-bit weights/activations,
(ii) requires no custom kernels or bespoke runtimes, and
(iii) runs unchanged on commodity CPUs, GPUs, and mainstream accelerators.

\subsection{Problem Setting and Requirements}
Edge inference imposes constraints that span models, toolchains, and hardware. Practical deployments require:
(1) \textbf{Cross-platform compatibility} across CPUs, GPUs, and production accelerators~\cite{meuser2024revisiting};
(2) \textbf{FP accuracy parity} (or better) on CNNs \emph{and} Transformers for realistic tasks~\cite{reddi2022mlperf};
(3) \textbf{Low engineering overhead} without re-implementing runtimes or kernels~\cite{nagel2021white}.
Additionally, robustness is increasingly a first-class requirement; training procedures should be compatible with adversarial defenses~\cite{li2024investigating}.

\subsection{Key Idea: Co-Design Across Training and Inference}
\FWName{} adopts a system-wide perspective that couples training dynamics with deployment constraints:
\begin{enumerate}
  \item \textbf{Training-time noise injection} tailored to sub-8-bit quantization, aligning weight/activation statistics with the discretization the hardware will realize.
  \item \textbf{Operator/format choices} restricted to widely available primitives, avoiding custom kernels and preserving framework/runtime compatibility.
  \item \textbf{Execution-aware calibration} that respects memory bandwidth and vectorization on commodity xPUs, so accuracy gains do not regress latency.
\end{enumerate}
This co-design, absent in prior work, is what enables \FWName{} to \textbf{improve both accuracy and end-to-end latency without special deployment support}.

\subsection{Why It Works}
The noise-injection loop shapes training to the quantized operating region, reducing post-training mismatch and activation outliers.
Because the induced noise is compatible with standard adversarial-training objectives (e.g., PGD-style perturbations), \FWName{} can be composed with robustness defenses in a straightforward manner~\cite{li2024investigating}.
Constrained operator choices preserve portability and keep the implementation faithful to production kernels.

\subsection{Evidence and Scope}
Across evaluated CNNs and Transformers, \FWName{} attains sub-8-bit W/A with parity or improvements over FP baselines, and does so using stock toolchains (no custom runtime).
The framework targets \emph{commodity} hardware and mainstream accelerators; portability is a design constraint, not a post-hoc optimization.
We focus on deployment-relevant tasks/datasets and report both accuracy and latency.

\subsection{Separation from Prior Work}
\textit{HAQ}~\cite{wang2019haq} optimizes mixed precision but assumes hardware support and is not designed to integrate with robustness techniques.
\textit{TQN}~\cite{fan2020training} introduces quantization noise during training yet exhibits accuracy drops and targets GPU-centric settings.
\textit{SMOL}~\cite{savarese2022not} emphasizes compression but lacks a realistic deployment path on commodity hardware. \textit{LSQ}~\cite{esser2019learned} leads to significant performance degradation in transformers \cite{xu2023q}. In contrast, \FWName{} achieves FP-parity (or better) at sub-8-bit W/A on both CNNs and Transformers without distillation, architecture changes, or custom runtimes, and runs unchanged on commodity xPUs.
In contrast, \FWName{} satisfies accuracy, portability, and engineering-effort criteria concurrently.

\paragraph{Takeaways.}
\FWName{}’s contribution is a deployment-first quantization method that (i) achieves sub-8-bit parity or gains on both CNNs and Transformers, (ii) requires no bespoke runtime support, and (iii) is portable across commodity xPUs.
\section{Complete \FWName{-QAT} Training Algorithm} \label{app:qat_algo}
\noindent\emph{Algorithm~\ref{alg:titan_qat} outlines \FWName{-QAT}, a two-phase procedure that first learns per-channel precisions via noise-driven discovery and then fine-tunes the discretized model under integer execution using a straight-through estimator (STE).}

\noindent\textbf{Description.}
\FWName{-QAT} proceeds in two phases.
We initialize a learnable precision palette \(v\) and per-channel logits \(z^{(l)}\).

\begin{itemize}
\item \emph{Phase I (noise-driven discovery):} we anneal the temperature \(\tau\), compute soft precision assignments \(a_{i,r}=\operatorname{softmax}_{r}(\tau z^{(l)}_{i,r})\), and form expected precisions \(s^{(l)}_{i}=\sum_{r}a_{i,r}v_r\).
Channels are permuted so equal precisions are contiguous (line~\ref{ln:permute}).
Group-wise weight and activation noises with amplitude \(\sigma(s^{(l)}_{i})\), scaled by per-group maxima, emulate quantization perturbations.
Training minimizes \(\mathcal{L}=L(w+\text{noise})+\lambda\sum_{l,i}\log_{2}\!\big(1+e^{-s^{(l)}_{i}}\big)\) while capping the palette to \(\le 8\) bits; we then discretize to \(p^{(l)}_{i}=1+\operatorname{round}\!\big(\log_{2}(1+e^{-s^{(l)}_{i}})\big)\).

\item \emph{Phase II (quantized fine-tuning):} we quantize \((w,x)\) per group using \((p,\text{GroupSize}_{\text{wt}},\text{GroupSize}_{\text{act}})\), execute integer MACs in the forward pass, backpropagate through \(\operatorname{round}(\cdot)\) via STE, and update master FP32 weights to convergence.
\end{itemize}

\begin{algorithm*}
\caption{\FWName{-QAT}: two-phase training with channel-wise noise and precision learning}
\label{alg:titan_qat}
% \footnotesize
\begin{algorithmic}[1]
  \Procedure{\FWName{-QAT}}{$k, w, \tau_{\text{final}}, \lambda, L,
                             \text{GroupSize}_{\text{act}},
                             \text{GroupSize}_{\text{wt}}$}

    % ----------  initialization  ----------
    \State Initialize palette $v\!\in\!\mathbb{R}^k$ and logits
           $z^{(l)}\!\in\!\mathbb{R}^{d_l\times k}$ \Comment{Lines 1–2}

    % ----------  phase I: noise-driven discovery  ----------
    \For{$t\gets1$ \textbf{to} $T_1$} \Comment{\textbf{Phase I}}
      \State $\tau \gets (\tau_{\text{final}})^{t/T_{1}}$ \Comment{Anneal temperature}
      \For{each layer $l$}
        \For{channel $i=1{:}d_l$} \Comment{Soft precision}
          \State $a_{i,r} \gets \operatorname{softmax}_{r}(\tau z^{(l)}_{i,r})$
          \State $s^{(l)}_{i} \gets \sum_{r} a_{i,r} v_r$
        \EndFor
        \State \textbf{Permute} channels s.t.\ equal precisions are contiguous
                                                   \label{ln:permute}
        \For{channel $i$}           \Comment{Weight noise (group-wise)}
          \State $g \gets \lfloor i/\text{GroupSize}_{\text{wt}}\rfloor$
          \State $w_{\max}^{(l,g)}\gets\max_{j\in g}|w^{(l)}_j|$
          \State Sample $\varepsilon^{(t)}_{w,i}\!\sim\!\mathcal{U}(\pm1)$
          \State $w^{(l)}_{i}\gets w^{(l)}_{i}
                 +\sigma(s^{(l)}_{i})\varepsilon^{(t)}_{w,i}w_{\max}^{(l,g)}$
        \EndFor
        \For{channel $i$}           \Comment{Activation noise (group-wise)}
          \State $g \gets \lfloor i/\text{GroupSize}_{\text{act}}\rfloor$
          \State $x_{\max}^{(l,g)}\gets\max_{j\in g}|x^{(l)}_j|$
          \State Sample $\varepsilon^{(t)}_{x,i}\!\sim\!\mathcal{U}(\pm1)$
          \State $x^{(l)}_{i}\gets x^{(l)}_{i}
                 +\frac{\sigma(s^{(l)}_{i})}{2}\varepsilon^{(t)}_{x,i}x_{\max}^{(l,g)}$
        \EndFor
      \EndFor
      \State $\mathcal{L}=L(w+\text{noise})+\lambda\sum_{l,i}\log_{2}(1+e^{-s^{(l)}_{i}})$
      \State Update $w,v,z$ via back-prop; clip $v$ to make sure precision $\leq 8$
    \EndFor
    \State
    % ----------  precision discretisation  ----------
    \For{each layer $l$, channel $i$}
      \State $p^{(l)}_{i}=1+\operatorname{round}\left(\log_{2}(1+e^{-s^{(l)}_{i}})\right)$
    \EndFor
    \State

    \For{$t \gets T_{1}+1$ \textbf{to} $T_{2}$} \Comment{\textbf{Phase II: Quantized fine-tuning}}
      \State \textbf{Quantize:} compute per-group scale factors and map
            $\left(w,x\right) \!\to\! \left(w_q,x_q\right)$ using
            $(p,\text{GroupSize}_{\text{wt}},\text{GroupSize}_{\text{act}})$
      \State \textbf{Forward:} evaluate loss
            $\mathcal{L}=L(w_q,x_q)$ under integer MAC execution
      \State \textbf{Backward (STE):} back-propagate through
            $\operatorname{round}(\cdot)$ with the straight-through estimator
      \State \textbf{Update:} apply optimizer to master FP32 weights $w$
    \EndFor

  \EndProcedure
\end{algorithmic}

\end{algorithm*}

\newpage
\section{Architecture Support for \FWName{} } \label{app:arch}
In this appendix, we introduce the MAC unit we design for support \FGMPNN{s}. Note that as discussed in \S\ref{sec:hardware_support}, numerous mainstream xPUs already have support for \FGMPNN{s}. We introduce our own architecture support for (1) evaluating inference latencies and (2) offering a design that is ready-to-ship for supporting \FGMPNN{s}.

\begin{figure}[b]
  \centering
  \includegraphics[width=0.9\linewidth]{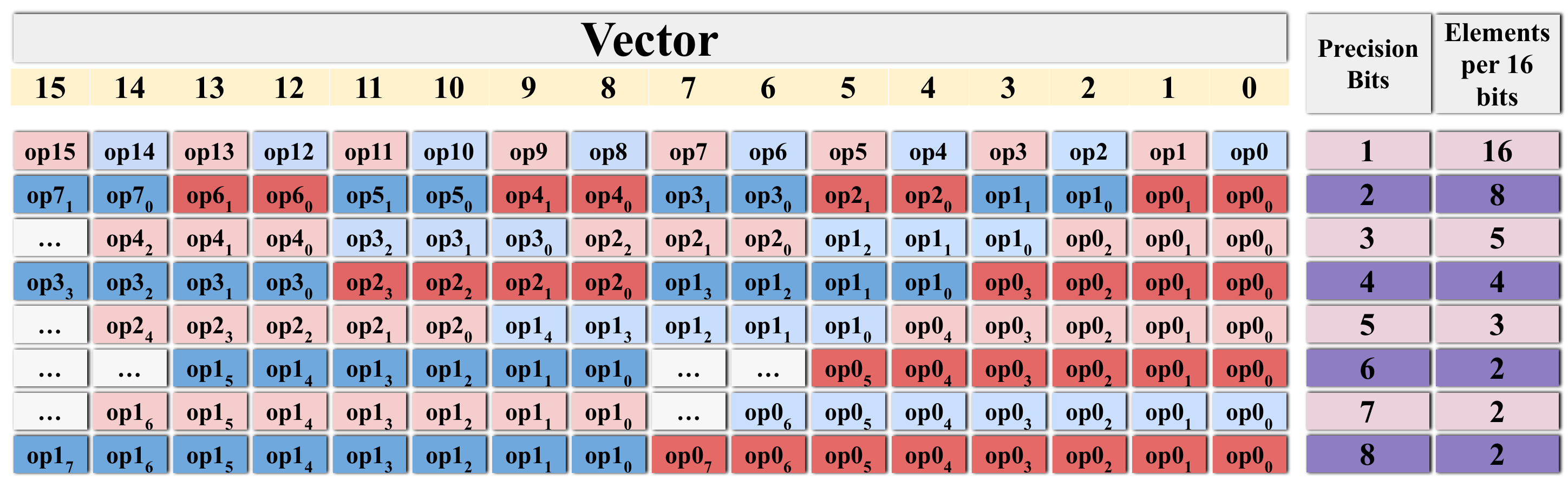}
  \caption{Operand packing scheme in \MACName. Consecutive rows (denoting
  different precisions) are coloured alternately.}
  \label{fig:operandpacking}
\end{figure}

\begin{figure*}
  \centering
  %
  %------------- left sub-figure ------------------------------------
  \begin{subfigure}{0.69\linewidth}
    \centering
    \includegraphics[width=\linewidth]{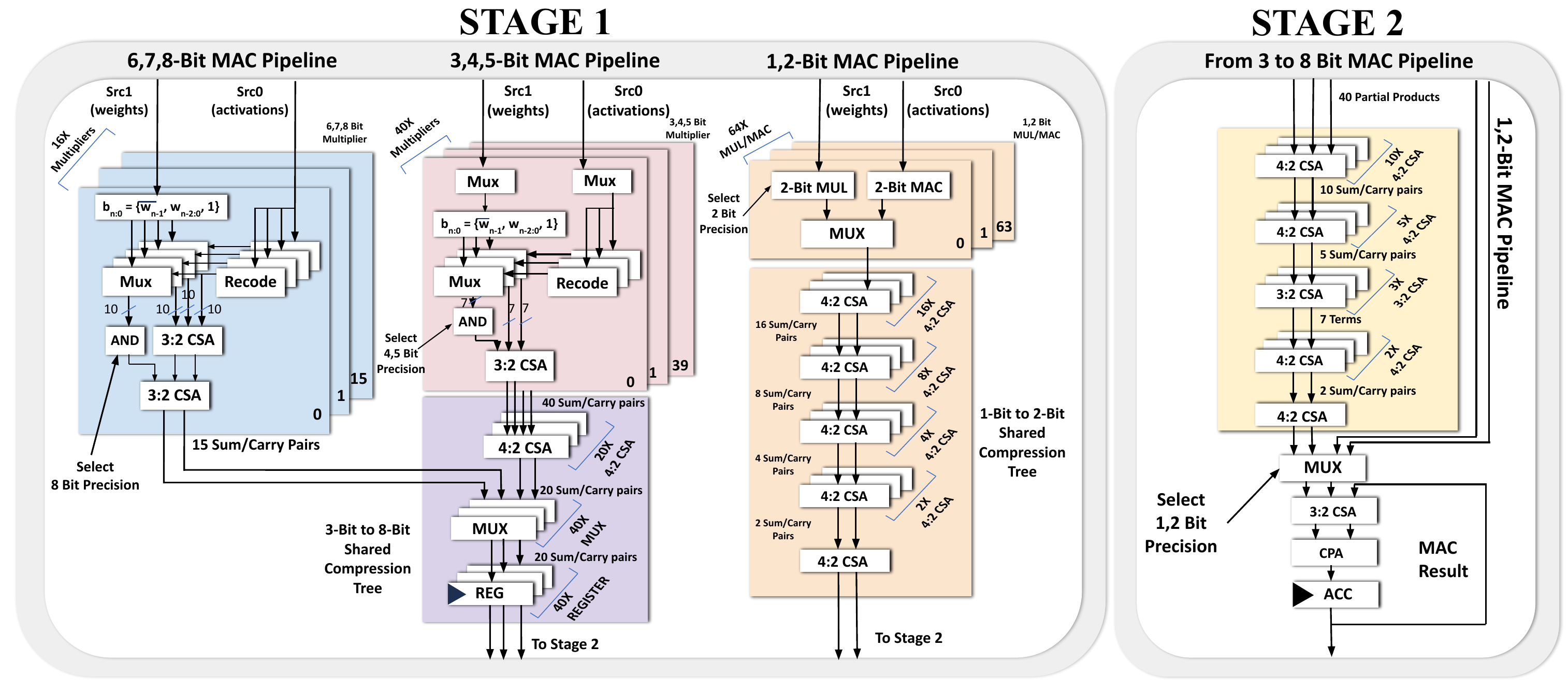}
    \caption{Configurable MAC design. CSA: carry-save adder; CPA: carry-propagate adder.}
    \label{fig:alu}
  \end{subfigure}
  \hfill
  %
  %------------- right sub-figure -----------------------------------
  \begin{subfigure}{0.29\linewidth}
    \centering
    \includegraphics[width=\linewidth]{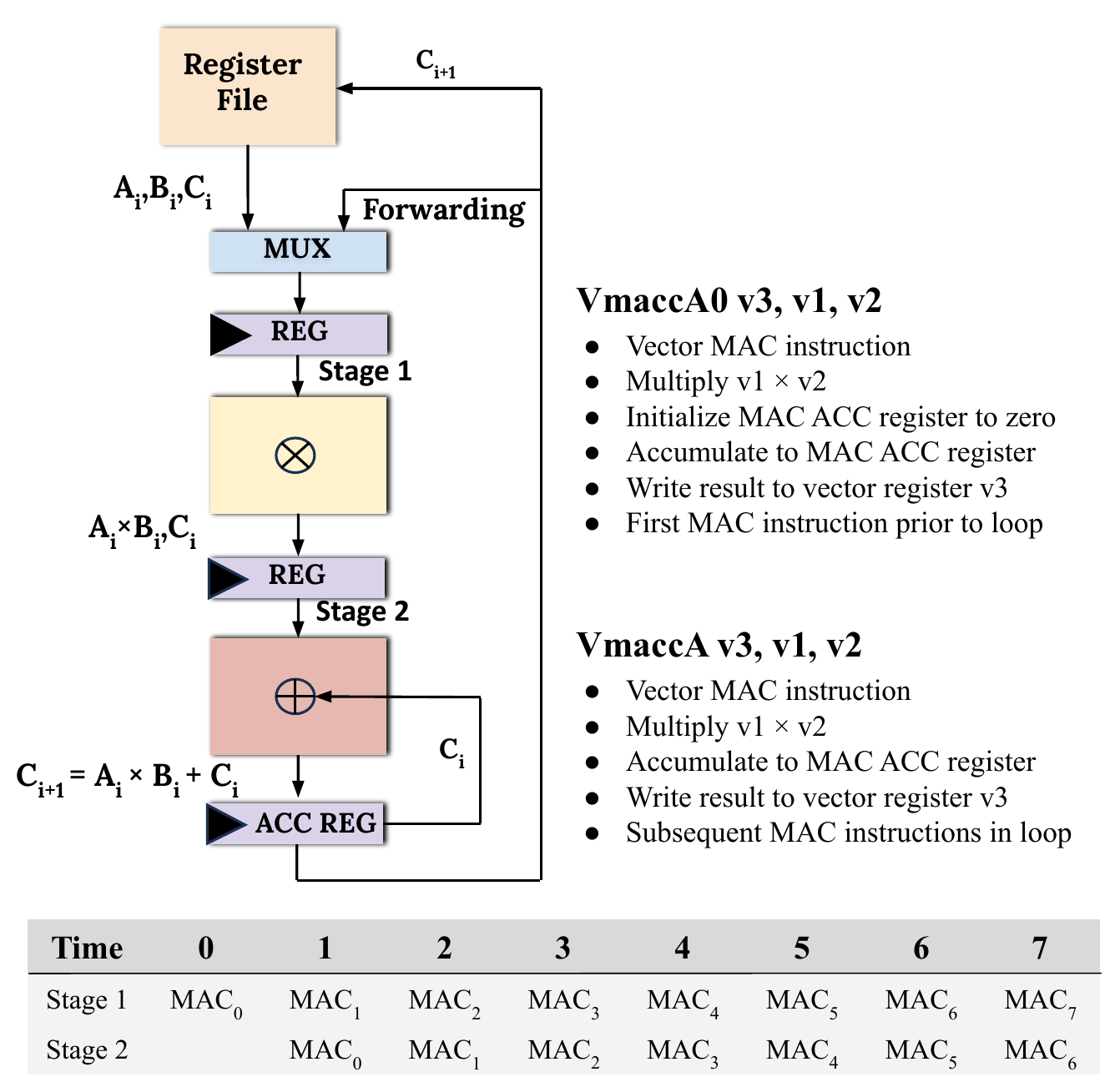}
    \caption{Partial-result accumulation in \MACName.}
    \label{fig:pipelining}
  \end{subfigure}
  \caption{\MACName{} datapath microarchitecture.  
           (\subref{fig:alu}) Configurable MAC unit;  
           (\subref{fig:pipelining}) pipeline-level partial-accumulation.}
  \label{fig:mac_pipeline}
\end{figure*}
\begin{figure*}
\centering
\includegraphics[width=0.33\textwidth]{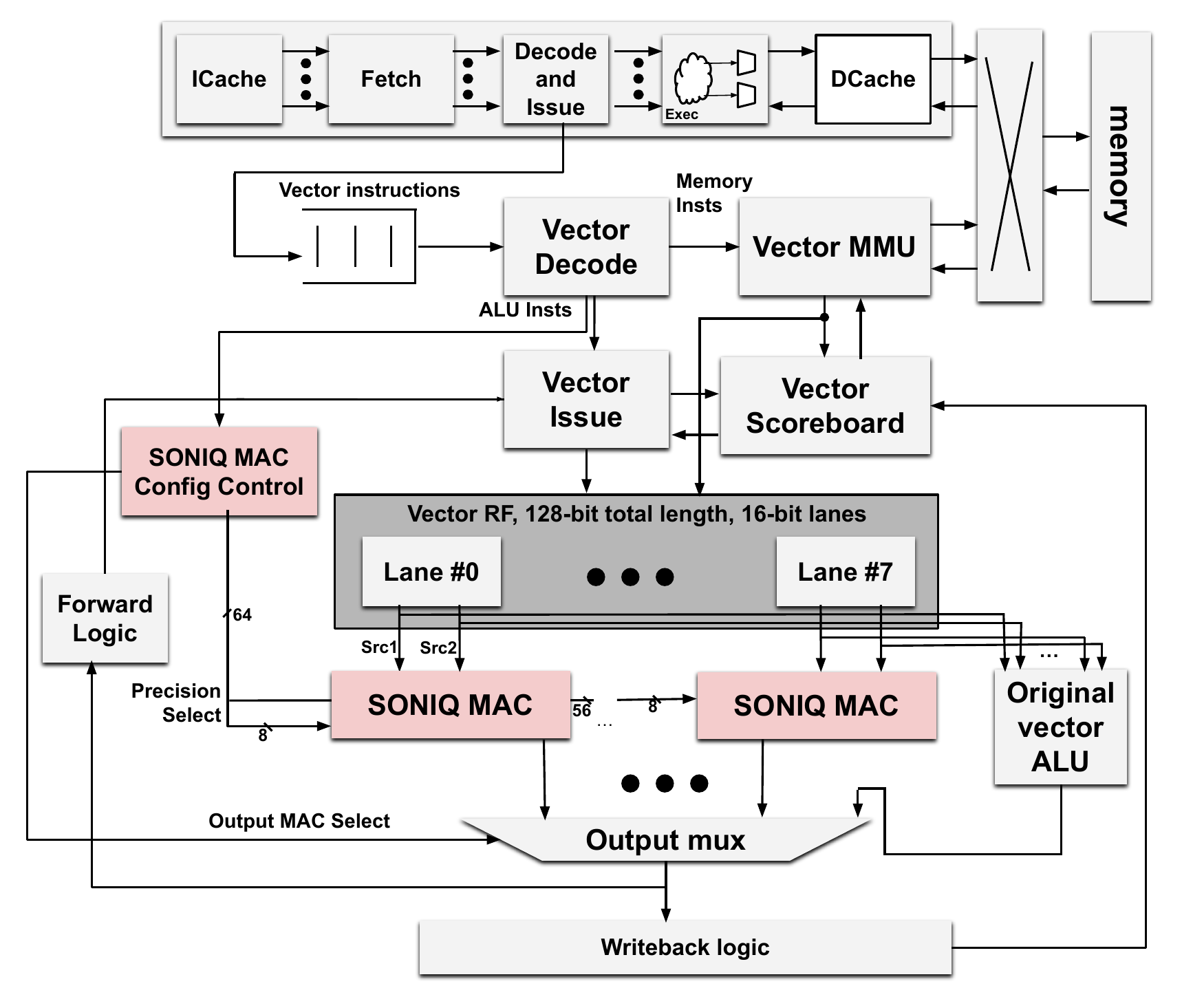}
\hfill
\includegraphics[width=0.63\textwidth]{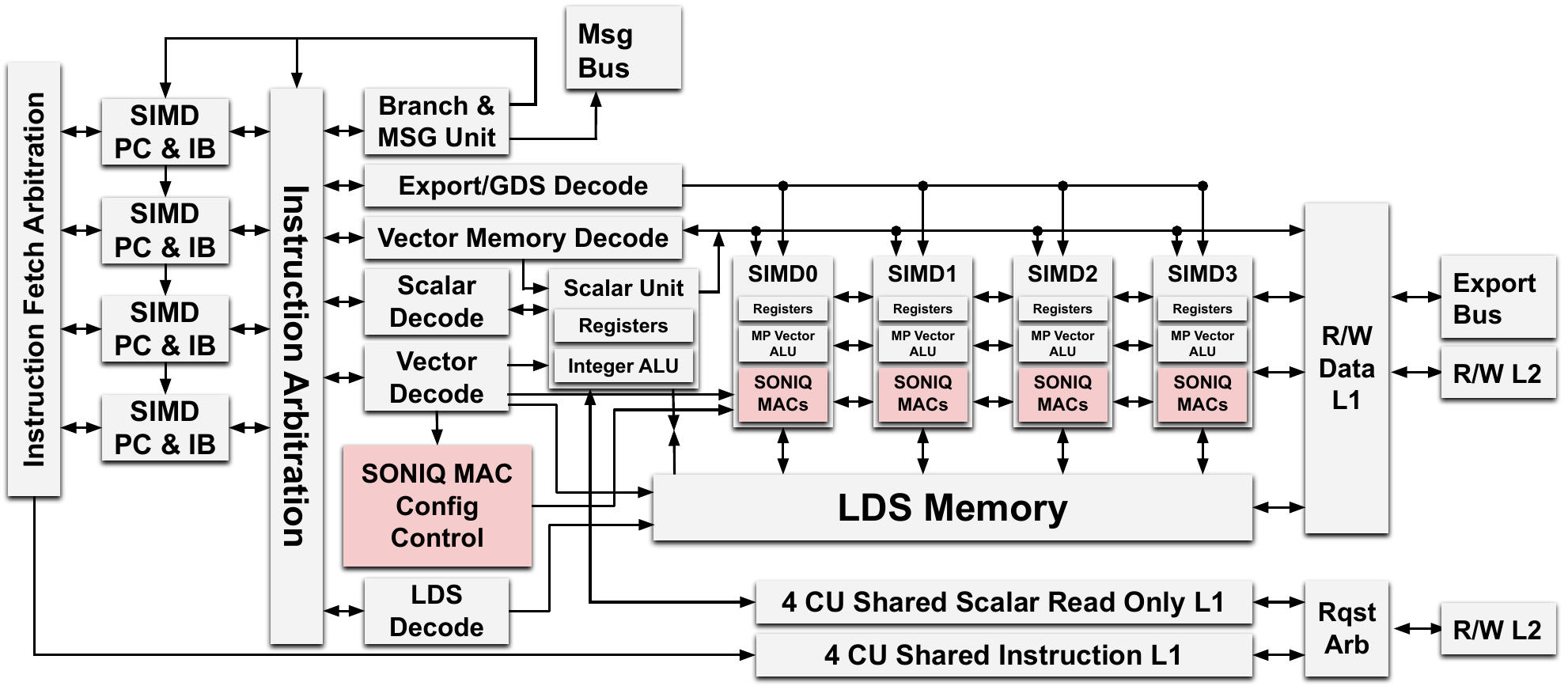}
\caption{CPU SIMD architecture (left) and base configurable GPU SIMD architecture (right) for \FGMPNN{s}. New blocks in pink; CPU drawing adopted from \cite{patsidis2020risc}, GPU drawing adopted from \cite{lec03_gpu_architectures}.}
\label{fig:cpuandgpusimd}
\end{figure*}

    \begin{figure*}
    \centering
    \includegraphics[width=0.7\linewidth]{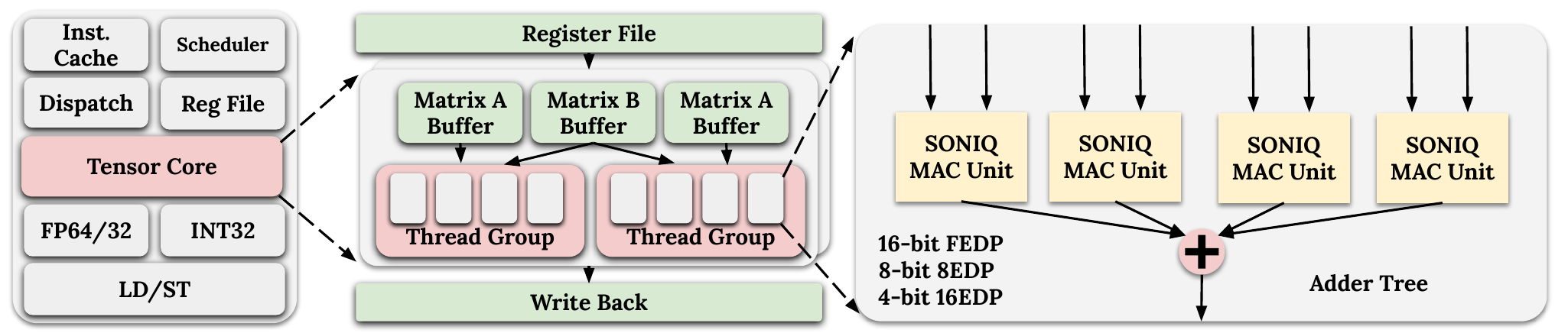}
    \caption{Applying \FWName{} MAC to Tensor Core; drawing adapted from \cite{guo2023olive}.}
    \label{fig:tensor_core}
    \end{figure*} 
\subsection{Configurable MAC Unit }
We present the design of a configurable MAC unit for inference on models trained with \FWName{-QAT}, referred to \textit{\FWName{} MAC} for short. It packs \FWName{} operands into 16-bit vector lanes, enabling 1–8 bit MAC operations while simplifying multiplexing and layout (Fig.~\ref{fig:operandpacking}).

As Fig.~\ref{fig:alu} illustrates, we partition multipliers into three banks—(i) two 1-bit MACs + one 2-bit MUL, (ii) 3–5-bit MULs, and (iii) 6–8-bit MULs—to achieve an optimized area-power-performance trade-off.

Each precision group uses a dedicated multiplier array, reducing switching
power. % by minimizing the product of partial-product width and Booth recoders per vector element. The \(3\text{–}5\)-bit unit handles  \(40\times3\)-bit products (5 per 16-bit lane),  \(32\times4\)-bit (4 per lane), or  \(24\times5\)-bit (3 per lane).  The \(6\text{–}8\)-bit unit processes \(16\) elements (2 per lane) at any of 6, 7, or 8 bits.   Because \(3\text{–}5\)-bit operands occupy varying bit positions inside a 16-bit lane, a lane MUX selects each element.  
Partial products are compressed by 3:2 trees—up to 40 and 16
\{\text{sum},\text{carry}\} pairs for the two units—and the \(3\text{–}5\)-bit
tree adds a 4:2 layer to halve 40 pairs to 20. Sixteen \{\text{sum},\text{carry}\} pairs (6–8 bit) and twenty pairs (3–5 bit)
feed a shared compressor to reduce area.

% At this point in the dataflow, the 16 \{sum,carry\} pairs from the 6,7,8-bit precision multipliers and the 20 \{sum,carry\} pairs from the 3,4,5-bit multipliers are multiplexed together into a common compression stage to reduce total area. The \{sum,carry\} pairs are flopped in pipeline registers to balance the number of logic levels per stage in 3-8-bit precisions vs 1-2 bit precisions.

 %Pipeline flops inserted here equalize logic depth between the 3–8 bit and 1–2 bit paths.

% Five additional levels of 4:2 and 3:2 compression are used to compress the 20 \{sum,carry\} pairs into a single \{sum,carry\} pair and is multiplexed with the single-cycle 1-2-bit precision \{sum,carry\} dataflow. The second stage pipeline register is used as a local accumulator to realize single-cycle throughput even with 2-cycle latency of the 3-8-bit precision MACs. The scheduler will insert a pipeline bubble if a single-cycle 1-2-bit precision MAC directly follows a 3-8-bit precision MAC.

Five extra 4{:}2/3{:}2 stages collapse the 20 \{\text{sum},\text{carry}\}
pairs to one, then multiplex it with the single-cycle 1–2-bit path.
The second-stage pipeline register acts as a local accumulator,
sustaining single-cycle throughput despite the 3–8-bit MAC’s two-cycle
latency; a scheduler bubble is inserted when a 1–2-bit MAC follows a
3–8-bit MAC.

Figure~\ref{fig:pipelining} shows a two-stage pipeline: multiplication completes in stage 1, accumulation in stage 2, with late multiplier gates occasionally slipping into stage 2 for timing closure.  

This MAC design requires only $\approx 5390$ NAND2 equivalent gate counts, a negligible quantity compared to the number of transistors on modern GPUs, such as the NVIDIA GA100 with 54.2 billion transistors \cite{ibrahim2012greda, nvidia_ampere}. The hardware costs of control logic required by \MACName{} are also trivial: only 1560 and 80 NAND2-equivalent gate counts are required to implement the control logic for vector unit integrations on CPUs and GPUs (see below), respectively.

%In the following, we illustrate how \MACName{} can be fit into mainstream architectures.
% \yl{the operand packing figure should be made larger; if space is an issue a lot of things about the MAC unit can be cut. }

% Requires \usepackage{subcaption}

\subsection{Integration into GPU Tensor Cores}

%\paragraph{Tensor Core baseline.}
%NVIDIA Tensor Cores accelerate matrix multiply–accumulate (MMA), the performance bottleneck of most deep-learning workloads, by fusing fixed-function MACs with on-core registers and a warp synchronous scheduler \cite{nvidia_ampere_whitepaper,markidis2018nvidia}.  In the A100 (Ampere) GPU, each of the $108$ Streaming Multiprocessors (SMs) integrates four Tensor Cores.  A single core sustains, per cycle--either \textbf{\textit{(1)}} a $16{\times}16 \times 16{\times}16$ FP16/TF32 MMA, \textbf{\textit{(2)}} a $16{\times}32 \times 32{\times}8$ INT8 MMA, or \textbf{\textit{(3)}} a $16{\times}32 \times 32{\times}8$ INT4 MMA--with accumulator promotion to FP32/INT32 \cite{nvidia_ampere, nvidia2023turing}.

%\paragraph{Enabling \FWName{} on Tensor Cores.}
\FWName{} employs a packed 16-bit data layout that existing Tensor Cores cannot
decode directly.  We therefore augment each  Tensor Core with
$256$ \FWName{} MAC units---mirroring the count of the resident FP16 ALUs.  A
scheduler bit selects, per warp, between the legacy FP16/INTx datapath and the
new \FWName{} datapath. The rest of the tensor core (register files, shared
memory, LSU) is unmodified.  Because each \FWName{} MAC consumes the
same operand bandwidth as an FP16 ALU while delivering two packed operations,
the modification \emph{doubles} the core’s effective throughput for
mixed-precision inference.

\vspace{-1em}
\subsection{Integration into CPU Vector Units}\label{subsec:cpu_vec}

%Vector units are ubiquitous in contemporary CPUs and GPUs, acceleratingelement-wise adds, multiplies, and activation functions essential to deep-learning workloads\cite{intel2023avx512,arm2023neon,intel2023sse}.

%\textbf{Baseline architecture.}
%Figure~\ref{fig:cpuandgpusimd} (left) sketches an out-of-order superscalar core---similar to \cite{patsidis2020risc}---whose scalar and 128-bit vector instructions share the \emph{fetch}, \emph{decode}, and \emph{issue} stages. After issue, vector instructions enter a dedicated SIMD pipeline.

%\textbf{Adding \FWName{} MACs.}
We integrate eight \FWName{} MAC units into
the SIMD pipeline of  an out-of-order superscalar core similar to \cite{patsidis2020risc}, as shown in Fig.~\ref{fig:cpuandgpusimd} (left).
A small control block (pink) broadcasts a 3-bit precision code to \FWName{} MAC,
selecting one of the eight supported precisions (1–8 bits). % These are the only new hardware blocks; all other core logic remains intact.  %Table~\ref{tab:gatecounts} shows the marginal area overhead.

%\textbf{ISA extensions.} 
In addition, two vector MAC instructions control the local accumulator illustrated in Figure \ref{fig:pipelining}:  \textbf{\textit{(1)}} \texttt{VMaccA0 v3,v1,v2} — multiplies \texttt{v1} and \texttt{v2}, clears the accumulator, and writes the product to both the accumulator and \texttt{v3}; used to start a MAC loop. \textbf{\textit{(2)}} \texttt{VMaccA v3,v1,v2} — multiplies \texttt{v1} and \texttt{v2}, adds the result to the current accumulator value, and stores the sum back to the accumulator and \texttt{v3}; used for all subsequent iterations.
% \begin{itemize}
% \item \lstinline{vmaccA0_Pn v3, v1, v2}  
%       \hfill(“accumulate and \emph{zero}”)\\
%       Multiplies \lstinline{v1} and \lstinline{v2}, clears the local accumulator,
%       and writes the product to both the accumulator and destination register
%       \lstinline{v3}.  Used to initialize an accumulation loop.
% \item \lstinline{vmaccA_Pn  v3, v1, v2}  
%       \hfill(“accumulate”)\\
%       Multiplies \lstinline{v1} and \lstinline{v2}, adds the result to the
%       current accumulator value, and writes the updated sum to the accumulator
%       and \lstinline{v3}.  Employed for all subsequent iterations.
% \end{itemize}

The suffix \lstinline{_Pn} encodes the common precision (\lstinline{n}=0--7 for
1–8-bit) for all lanes.  When channel groups of different precisions map to the
same register, we apply loop unrolling plus predicate masking on the final,
short vector to ensure homogeneous precision. Our experiments indicate that per-lane precision control offers negligible latency benefits, so we adopt the simpler broadcast scheme.

\subsection{Integration into GPU Vector Units}

The baseline GPU architecture in our case study is adopted from \cite{lec03_gpu_architectures}. The changes for supporting \ are minimal, as shown in Fig. \ref{fig:cpuandgpusimd} (right) in pink, which include: (1) integrating the \MACName{} into the SIMD units of each compute unit, and (2) including a new control module to select the precision configuration for each \MACName. GPUs already has native support for reduction sum operations across all MAC unit outputs. Two new instructions similar to \lstinline{vmaccA0_Pn} and \lstinline{vmaccA_Pn} for the CPU are introduced for the GPU, supporting MAC operations for precisions from 1 bit to 8 bits.

\end{document}